\documentclass{andp2012}
\usepackage[english]{babel}
\usepackage{graphics, bm, psfrag, amsmath, amssymb, epsfig, grffile, float, bbold}
\usepackage{subfigure}
\usepackage{dsfont}
\usepackage{color}
\usepackage{hyperref}
\usepackage{cuted}
\usepackage{widetext}

\title{The Role of Odd-Frequency Pairing in Multiband Superconductors}
\author[C. Triola]{Christopher Triola\inst{1,}\footnote{Corresponding author\quad E-mail:~\textsf{christopher.triola@physics.uu.se}}}
\author[J. Cayao]{Jorge Cayao\inst{1}}
\author[A.M. Black-Schaffer]{Annica M. Black-Schaffer\inst{1}}
\address[1]{Department of Physics and Astronomy, Uppsala University, Box 516, S-751 20 Uppsala, Sweden}
\shortauthors{C. Triola et al.}
\begin{abstract}
In this article we review recent progress in the understanding of multiband superconductivity and its relationship to odd-frequency pairing. We begin our discussion by reviewing the emergence of odd-frequency pairing in a simple two-band model, providing a brief pedagogical overview of the formalism. We then examine several examples of multiband superconducting systems in each case describing, both, the origin of the band degree of freedom and the nature of the odd-frequency pairing. Throughout, we attempt to convey a unified picture of how odd-frequency pairing emerges in these materials and propose that similar mechanisms are responsible for odd-frequency pairing in several analogous systems: layered two-dimensional heterostructures, double quantum dots, double nanowires, Josephson junctions, and systems described by isolated valleys in momentum space. We also review experimental probes of odd-frequency pairing in multiband systems, focusing on hybridization gaps in the electronic density of states, paramagnetic Meissner effect, and Kerr effect.   
\end{abstract}
\shortabstract
\begin{document}
\maketitle

\section{Introduction}

It is well-established that the symmetries of a superconducting order parameter influence many of the physical properties of a superconductor\cite{anderson1959theory,ferrell1959knight,anderson1959knight,sigrist1991phenomenological,van1995phase,tsuei2000pairing,balatsky2006impurity,qi2011topological}, including robustness to impurities\cite{anderson1959theory,balatsky2006impurity}, Knight shift as measured by nuclear magnetic resonance\cite{ferrell1959knight,anderson1959knight}, anisotropy in phase-sensitive measurements\cite{van1995phase}, and topological properties\cite{qi2011topological}. Within the standard BCS theory of superconductivity the order parameter, or mean field, usually denoted $\Delta$, is given in terms of equal-time expectation values of the form $\Delta\sim \langle \psi(t) \psi(t)\rangle$, which can be viewed as a many-body wavefunction describing pairs of electrons at the same time, $t$. Since electrons are fermions, this wavefunction must be antisymmetric under the simultaneous permutation of all quantum numbers describing the electrons, i.e.~spin and position degrees of freedom, for single band superconductors. This implies that, for superconductors with a single relevant band, order parameters with even spatial parity (like $s$- or $d$-wave) must correspond to a spin-singlet configuration, while order parameters with odd spatial parity ($p$- or $f$-waves) must correspond to spin-triplet states. 

A more accurate description of conventional superconductivity, however, starts with a retarded phonon-mediated interaction, described by Eliashberg theory\cite{eliashberg1960interactions,abrikosov2012methods,mahan2013many} and its generalizations\cite{scalapino1966strong,berk1966effect}. In this formalism the superconducting mean field is related to time-ordered expectation values, $\Delta(t-t')\sim \langle T \psi(t) \psi(t')\rangle$, and therefore necessarily depends on the relative time, $t-t'$, or, equivalently, the relative frequency, $\omega$. As a consequence, we obtain a symmetry constraint for this time-ordered expectation value of Cooper pairs, also known as the pair correlator or anomalous Green's function. As Berezinskii showed in 1974\cite{Berezinskii1974}, this allows for the possibility of odd-frequency (odd-$\omega$) order parameters which possess the opposite relationship between the spatial parity and spin configurations, i.e.~even-parity spin-triplet and odd-parity spin-singlet. We stress that this does not imply the breaking of time-reversal symmetry, since odd-$\omega$ order parameters are simply odd functions of the relative time coordinate, while the time-reversal operation includes complex conjugation\cite{kuzmanovski2017multiple,geilhufe2018symmetry,linder2017odd}.  

While Berezinskii's original proposal was made in the context of superfluid $^3$He, later works generalized the possibility of odd-$\omega$ order parameters to superconductivity\cite{kirkpatrick_1991_prl, belitz_1992_prb, BalatskyPRB1992, coleman_1993_prl, coleman_1994_prb, coleman_1995_prl}. However, since these original proposals, the thermodynamic stability of intrinsically odd-$\omega$ superconductors have been called into question\cite{heid1995thermodynamic, belitz_1999_prb, solenov2009thermodynamical, kusunose2011puzzle, FominovPRB2015}. Still, while the existence of such intrinsic odd-$\omega$ order parameters remains an intriguing theoretical question, a great deal of progress has been made studying the emergence of odd-$\omega$ pair correlations in systems with conventional equal-time order parameters\cite{bergeret2005odd,linder2017odd}. This latter possibility relies on the conversion of intrinsic even-$\omega$ superconducting correlations to odd-$\omega$ correlations, with a number of proposals in the literature realizing such symmetry conversion through a variety of different mechanisms\cite{BergeretPRL2001, bergeret2005odd, halterman2007odd, yokoyama2007manifestation, houzet2008ferromagnetic, EschrigNat2008, LinderPRB2008, crepin2015odd, YokoyamaPRB2012, Black-SchafferPRB2012, Black-SchafferPRB2013, TriolaPRB2014, tanaka2007theory, TanakaPRB2007,cayao2017odd, cayao2018odd, LinderPRL2009, LinderPRB2010_2, TanakaJPSJ2012, triola2016prl, triolaprb2016, black2013odd, sothmann2014unconventional, parhizgar_2014_prb, asano2015odd, komendova2015experimentally, burset2016all, komendova2017odd, kuzmanovski2017multiple, triola2017pair,keidel2018tunable, triola2018odd,fleckenstein2018conductance,asano2018green,triola2018oddnw}.

The prototypical example is the superconductor-ferromagnet (SF) junction, in which numerous theoretical works have demonstrated that the breaking of spin-rotational symmetry can convert conventional $s$-wave spin-singlet Cooper pairs to odd-$\omega$ spin-triplet pairs\cite{BergeretPRL2001, bergeret2005odd, halterman2007odd, yokoyama2007manifestation, houzet2008ferromagnetic, EschrigNat2008, LinderPRB2008, crepin2015odd}. Furthermore, experiments on these junctions have observed multiple signatures of the odd-$\omega$ spin-triplet pair correlations\cite{petrashov1994conductivity,giroud1998superconducting,petrashov1999giant,aumentado2001mesoscopic,zhu2010angular,di2015signature,di2015intrinsic}. Interestingly, it has also been demonstrated that odd-parity odd-$\omega$ pairing can emerge at the interface between a conventional even-parity superconductor and a normal metal (SN junction) due to broken spatial translation symmetry\cite{tanaka2007theory,TanakaPRB2007}. The magnitudes of the odd-$\omega$ correlations have been shown to dominate over the even-$\omega$ amplitudes at discrete energy levels coinciding exactly with peaks in the local density of states (LDOS)\cite{TanakaPRB2007}, establishing a relationship between odd-$\omega$ pairing and McMillan-Rowell oscillations\cite{rowell1966electron,rowell1973tunneling} as well as midgap Andreev resonances\cite{alff1997spatially,covington1997observation,wei1998directional}.

In this article we focus on several recent works exploring the intriguing possibility of realizing odd-$\omega$ pair correlations in multiband superconductors, without the need for magnetism or interfaces. In 2013 it was shown that odd-$\omega$ pairing should arise ubiquitously in such systems due to the presence of interband hybridization\cite{black2013odd}. Since this interband hybridization is generally uniform throughout the bulk of a multiband superconductor, this establishes the existence of a class of systems that should host bulk odd-$\omega$ pairing. Since the original proposal, multiple additional works have built on this concept\cite{ sothmann2014unconventional, parhizgar_2014_prb,asano2015odd,  komendova2015experimentally,triola2016prl,triolaprb2016, burset2016all, ebisu2016theory, komendova2017odd, kuzmanovski2017multiple, triola2017pair, balatsky2018odd, keidel2018tunable, triola2018odd,fleckenstein2018conductance,asano2018green,triola2018oddnw} generalizing the original idea to related systems, focusing on specific physical examples, or studying the experimental consequences of odd-$\omega$ pairing in these multiband systems. With many known multiband superconductors with highly unconventional features, such as Sr$_2$RuO$_4$\cite{maeno1994superconductivity,maeno2012}, iron-based superconductors \cite{hunte2008two, kamihara2008iron, ishida2009extent, cvetkovic2009multiband, stewart2011superconductivity}, MgB$_2$ \cite{nagamatsu2001superconductivity, bouquet2001specific, brinkman2002multiband, golubov2002specific,iavarone2002two}, and UPt$_3$\cite{stewart_prl_1984, adenwalla1990,sauls1994,strand_prl_2009,strand_science_2010} it remains a very interesting question how much odd-$\omega$ superconductivity contributes to the physical properties of these and related systems.

It is important to note that the presence of additional band degrees of freedom in these multiband superconductors leads, not only to novel methods of inducing odd-$\omega$ pairing, but also to a broader classification scheme for the symmetry of the Cooper pairs. For a generic superconducting system with multiple bands crossing the Fermi level, we write the anomalous Green's function as the time-ordered expectation value: $F(1,2)=-\langle T \psi_{\sigma_1,x_1,\alpha_1}(t_1)\psi_{\sigma_2,x_2,\alpha_2}(t_2) \rangle$, where $\sigma_i$, $x_i$, $\alpha_i$, and $t_i$ represent the spin, positions, band, and time degrees of freedom. We emphasize that this band index has its origin in the electronic degrees of freedom of the superconductor and could stem from any of the indices characterizing the system, including the atomic orbital, sublattice, layer, dot, lead, or valley indices, as we discuss in Sec.~\ref{sec:other}. Accounting for the symmetry under the exchange of each of these pairs of indices we necessarily have $F(1,2)=-F(2,1)$. This leads to eight possible symmetry classes\cite{black2013odd, asano2015odd, triolaprb2016, linder2017odd}, four even-$\omega$ and four odd-$\omega$ classes as seen in Table \ref{table:classification}. 

\begin{table}
\begin{tabular}{c || c | c | c | c || c | c | c | c |}
        & 1 & 2 & 3 & 4 & 5 & 6 & 7 & 8 \\
\hline 
Spin ($\mathcal{S}$ )   & - & + & + & - & + & - & - & +  \\ 
\hline
Parity ($\mathcal{P}$)  & + & - & + & - & + & - & + & -  \\
\hline
Orbital ($\mathcal{O}$) & + & + & - & - & + & + & - & -  \\
\hline
 Time  ($\mathcal{T}$)  & + & + & + & + & - & - & - & -  \\
\end{tabular}
\caption{Characterization of the eight symmetry classes for superconducting pair amplitudes allowed by Fermi-Dirac statistics. Each column represents a different symmetry class, with the sign, $\pm$, representing the symmetry of the anomalous Green's function under the exchange of the index indicated in the far left column: $\mathcal{S}F_{\sigma_1,\sigma_2}= F_{\sigma_2,\sigma_1}$ (spin); $\mathcal{P}F_{x_1,x_2}=F_{x_2,x_1}$ (parity); $\mathcal{O}F_{\alpha_1,\alpha_2}=F_{\alpha_2,\alpha_1}$ (band or similar); and $\mathcal{T}F_{t_1,t_2}=F_{t_2,t_1}$ (time).}
\label{table:classification}
\end{table}

In the remainder of this article we provide, in Sec. \ref{sec:overview}, a pedagogical overview of the emergence of odd-$\omega$ pairing in a simple two-band model due to interband hybridization, as well as a discussion of when we should expect to find odd-$\omega$ pairing in a generic superconducting system. Following the overview, we focus in Sec.~\ref{sec:examples} on specific proposals for realizing odd-$\omega$ pairing in well-known multiband superconductors, and also in materials and systems possessing related active electronic degrees of freedom, such as layer, dot, lead or valley indices. In Sec.~\ref{sec:experiment} we discuss proposed experimental signatures of odd-$\omega$ pairing in multiband superconductors. Finally, in Sec.~\ref{sec:conclusions} we conclude our discussion.

\section{General Results for Multiband Odd-frequency Pairing}
\label{sec:overview}

\subsection{Two Band Model with Interband Hybridization}
\label{sec:twoband_criteria}
To illustrate the emergence of odd-$\omega$ pairing in multiband superconductors we study the pair amplitudes associated with the following two band Hamiltonian\cite{black2013odd}:
\begin{equation}
H=\frac{1}{2}\sum_{\textbf{k}} \Psi^\dagger_{\textbf{k}} \left( \begin{array}{cc} \hat{h}_{\textbf{k}} & \hat{\Delta}_{\textbf{k}} \\
\hat{\Delta}^\dagger_{\textbf{k}} & -\hat{h}^*_{-\textbf{k}} \end{array} \right) \Psi_{\textbf{k}}, 
\label{eq:ham2band_compact}
\end{equation}
using the basis:
\begin{equation}
\Psi^\dagger_{\textbf{k}}= \left( 
c^\dagger_{\uparrow,1,\textbf{k}} 
c^\dagger_{\uparrow,2,\textbf{k}}
c^\dagger_{\downarrow,1,\textbf{k}}
c^\dagger_{\downarrow,2,\textbf{k}}
c_{\uparrow,1,-\textbf{k}}
c_{\uparrow,2,-\textbf{k}}
c_{\downarrow,1,-\textbf{k}} 
c_{\downarrow,2,-\textbf{k}}  
\right),
\end{equation}
where $c^{\dagger}_{\sigma,\alpha,\textbf{k}}$ ($c_{\sigma,\alpha,\textbf{k}}$) creates (annihilates) a fermionic quasiparticle with spin $\sigma$ in band $\alpha$ and with momentum $\textbf{k}$, together with the definitions:
\begin{equation}
\begin{aligned}
\hat{h}_{\textbf{k}}=\left(\begin{array}{cc}
\xi_{1,\textbf{k}} & \Gamma  \\
\Gamma^* & \xi_{2,\textbf{k}}  
\end{array} \right) \otimes \hat{\sigma}_0, \
\hat{\Delta}_{\textbf{k}}=\left(\begin{array}{cc}
\Delta_{1,\textbf{k}} & 0 \\
 0 & \Delta_{2,\textbf{k}} 
\end{array} \right)\otimes i\hat{\sigma}_2,
\end{aligned}
\label{eq:ham2band_compact_definitions}
\end{equation}
where $\hat{\sigma}_0$ and $\hat{\sigma}_{i=1,2,3}$ are the identity and Pauli matrices in spin space. Here $\xi_{\alpha,\textbf{k}}$ is the energy dispersion of band $\alpha$, $\Gamma$ is a measure of the interband hybridization, and $\Delta_{\alpha,\textbf{k}}$ is the superconducting order parameter in band $\alpha$, here assumed for simplicity to be spin-singlet in nature, although it is trivial to extend the derivation to spin-triplet pairing. We note that this kind of interband hybridization, $\Gamma$, is intrinsic to a superconductor whenever there is a mismatch between the the quasiparticles of the normal state and the orbital character of the Cooper pairs or, alternatively, it can arise from scattering processes in the presence of disorder\cite{black2013odd,komendova2015experimentally,komendova2017odd}. Hence, for generic multiband superconductors we expect it to be nonzero. 

To study the pair amplitudes associated with Eq.~\eqref{eq:ham2band_compact}, it is convenient to define the Nambu-Gorkov Green's functions as follows\cite{abrikosov2012methods,mahan2013many}:
\begin{equation}
\begin{aligned}
G_{\sigma,\alpha;\sigma',\alpha'}(\textbf{k},\tau)&=-\langle T_\tau c_{\sigma,\alpha,\textbf{k}}(\tau) c^\dagger_{\sigma',\alpha',\textbf{k}}(0) \rangle, \\
F_{\sigma,\alpha;\sigma',\alpha'}(\textbf{k},\tau)&=-\langle T_\tau c_{\sigma,\alpha,-\textbf{k}}(\tau) c_{\sigma',\alpha',\textbf{k}}(0) \rangle, \\
\bar{G}_{\sigma,\alpha;\sigma',\alpha'}(\textbf{k},\tau)&=-\langle T_\tau c^\dagger_{\sigma,\alpha,-\textbf{k}}(\tau) c_{\sigma',\alpha',-\textbf{k}}(0) \rangle, \\
\bar{F}_{\sigma,\alpha;\sigma',\alpha'}(\textbf{k},\tau)&=-\langle T_\tau c^\dagger_{\sigma,\alpha,\textbf{k}}(\tau) c^\dagger_{\sigma',\alpha',-\textbf{k}}(0) \rangle, 
\end{aligned}
\label{eq:greens_definition}
\end{equation}
where $\tau$ is imaginary time and $T_\tau$ is the $\tau$-ordering operator. With these definitions, it is straightforward to derive the following equations of motion:
\begin{equation}
\left(\begin{array}{cc}
i\omega_n-\hat{h}_{\textbf{k}} & -\hat{\Delta}_{\textbf{k}} \\
-\hat{\Delta}^\dagger_{\textbf{k}} & i\omega_n+\hat{h}_{-\textbf{k}}^* 
\end{array} \right) \left(\begin{array}{cc}
\hat{G}(\textbf{k},i\omega_n) & \hat{F}(\textbf{k},i\omega_n) \\
\hat{\bar{F}}(\textbf{k},i\omega_n) & \hat{\bar{G}}(\textbf{k},i\omega_n)
\end{array} \right) =\mathbb{1},
\label{eq:2band_eom}
\end{equation} 
where we have Fourier-transformed the Green's functions from imaginary time, $\tau$, to Matsubara frequency, $i\omega_n$, and $\mathbb{1}$ is the 8$\times$8 identity matrix in band $\times$ spin $\times$ particle-hole space. For simplicity, we assume time-reversal symmetry such that $\xi_{\alpha,-\textbf{k}}=\xi_{\alpha,\textbf{k}}$ and for the moment we also set $\Gamma=\Gamma^*$.

After some straightforward algebra we find that the anomalous Green's function, $\hat{F}$, is given by:
\begin{widetext}
\begin{equation}
\begin{aligned}
\hat{F}(\textbf{k},i\omega_n)=&\frac{1}{D_{\textbf{k},i\omega_n}}\left(\begin{array}{cc}
\Delta_{1,\textbf{k}}\left[(i\omega)^2-E_{2,\textbf{k}}^2 \right]-\Delta_{2,\textbf{k}}\Gamma^2 & \Gamma\left[-i\omega_n\left(\Delta_{1,\textbf{k}}-\Delta_{2,\textbf{k}} \right)+\Delta_{1,\textbf{k}}\xi_{2,\textbf{k}}+\Delta_{2,\textbf{k}}\xi_{1,\textbf{k}} \right] \\
\Gamma\left[i\omega_n\left(\Delta_{1,\textbf{k}}-\Delta_{2,\textbf{k}} \right)+\Delta_{1,\textbf{k}}\xi_{2,\textbf{k}}+\Delta_{2,\textbf{k}}\xi_{1,\textbf{k}} \right] & \Delta_{2,\textbf{k}}\left[(i\omega)^2-E_{1,\textbf{k}}^2 \right]-\Delta_{1,\textbf{k}}\Gamma^2
\end{array} \right) \otimes i\hat{\sigma}_2,
\end{aligned}
\label{eq:2band_anomalous}
\end{equation}
\end{widetext}
where we define:
\begin{equation}
\begin{aligned}
D_{\textbf{k},i\omega_n}& =(i\omega_n)^4 -(i\omega_n)^2\left[ E_{1,\textbf{k}}^2+E_{2,\textbf{k}}^2+2\Gamma^2 \right]+E_{1,\textbf{k}}^2E_{2,\textbf{k}}^2 \\
&+ \Gamma^2\left(\Delta_{1,\textbf{k}}\Delta_{2,\textbf{k}}^*+\Delta_{1,\textbf{k}}^*\Delta_{2,\textbf{k}}+\Gamma^2-2\xi_{1,\textbf{k}}\xi_{2,\textbf{k}} \right), \\
E_{\alpha,\textbf{k}}&=\sqrt{\xi_{\alpha,\textbf{k}}^2 + \Delta_{\alpha,\textbf{k}}^2}.
\end{aligned}
\label{eq:denomenator}
\end{equation}
From Eq. (\ref{eq:2band_anomalous}) we directly see that the pair amplitude's spin structure is given by $i\hat{\sigma}_2$, therefore the spin-singlet nature of the Cooper pairs remains completely unaffected by the presence of the interband hybridization. 

Inspecting the intraband pairing, given by the diagonal elements of the matrix in Eq.~(\ref{eq:2band_anomalous}), we find all amplitudes being even in Matsubara frequency and spatial parity, corresponding to pair amplitudes in the first column of Table \ref{table:classification}. However, turning our attention to the interband pairing, given by the off-diagonal elements of the matrix in Eq. (\ref{eq:2band_anomalous}), we find that these pair amplitudes have both even- and odd-$\omega$ terms. Notably, we see that the even-$\omega$ amplitude is also even in the band index, and, thus, also belongs to the symmetry class in column 1 of Table \ref{table:classification}, while the odd-$\omega$ amplitude is odd in the band index and, thus, belongs to the symmetry class in column 7 of Table \ref{table:classification}. As a consequence, for this model, all pair amplitudes that are even in the band index are also even in frequency while the odd-band pairing is entirely odd-$\omega$\cite{black2013odd,komendova2015experimentally,komendova2017odd}. This complete reciprocity between frequency and band parity also holds for more complicated models as long as the order parameter appearing in the Hamiltonian is even in the band index and no other symmetries are broken.    

If we relax the assumption that the interband hybridization is real, instead setting $\Gamma=|\Gamma|e^{i\phi}$, we find that the odd-$\omega$ pair amplitude is given by\cite{asano2015odd}:
\begin{equation}
F_{odd}(\textbf{k};i\omega_n)=\frac{i\omega_n |\Gamma|}{D'_{\textbf{k},i\omega_n}}\left(\Delta_{1,\textbf{k}}e^{-i\phi}-\Delta_{2,\textbf{k}}e^{i\phi}\right)\hat{\rho}_2\otimes\hat{\sigma}_2, 
\label{eq:odd_complex_gamma}
\end{equation}
where $\hat{\rho}_2$ is a Pauli matrix in band space and $D'_{\textbf{k},i\omega_n}$ is still an even function of $\textbf{k}$ and $i\omega_n$. From Eq. (\ref{eq:odd_complex_gamma}) we see that, when $\phi=0$ (real $\Gamma$) the odd-$\omega$ pairing in this two-band model is proportional to $i\omega_n\Gamma\left(\Delta_{1,\textbf{k}}-\Delta_{2,\textbf{k}} \right)$, and is therefore non-zero whenever there is both interband hybridization, $\Gamma\neq 0$, and a difference between the two gaps, $\Delta_{1,\textbf{k}}-\Delta_{2,\textbf{k}}\neq 0$, which is usually the case in multiband superconductors. When $\phi=\tfrac{\pi}{2}$ (imaginary $\Gamma$) the odd-$\omega$ pairing is instead proportional to $i\omega_n\Gamma\left(\Delta_{1,\textbf{k}}+\Delta_{2,\textbf{k}} \right)$ and is therefore non-zero as long as $\Gamma\neq 0$ and $\Delta_{1,\textbf{k}}\neq-\Delta_{2,\textbf{k}}$. Between these two extremes we find a non-zero odd-$\omega$ interband pair amplitude regardless of the values of the two gaps, as long as there is finite interband hybridization. Moreover, given that this kind of interband hybridization should be present in most multiband superconductors, we expect odd-$\omega$ pairing to be ubiquitous in multiband superconductors\cite{black2013odd,komendova2015experimentally,komendova2017odd}.

\subsection{Generalization to Arbitrary Hamiltonians}
\label{sec:gen}
To understand how the results for the simple two-band model generalize to more complicated models, it is instructive to consider a generic model:
\begin{equation}
\begin{aligned}
\label{eq:hamgenband_compact}
H=&\sum_{n,m} \left( \begin{array}{cc} 
c^\dagger_n & c_n \end{array} \right) \left( \begin{array}{cc} 
h_{nm} & \Delta_{nm} \\
\Delta^\dagger_{nm} & -h_{nm}^* \end{array} \right) \left( \begin{array}{c} 
c_m \\
c^\dagger_m \end{array} \right),
\end{aligned}
\end{equation}
where the indices $n,m$ label all degrees of freedom for the quasiparticles, including: spin, position, band/orbital, sublattice, etc. It is easy to see that the Hamiltonian in Eq. (\ref{eq:ham2band_compact}) is a particular example of a Hamiltonian of this form. Moreover, any Hermitian Hamiltonian with a BCS-like order parameter may be written in this form. 

To examine the pair amplitudes for the Hamiltonian in Eq.~\eqref{eq:hamgenband_compact} we use Green's functions which are merely generalized versions of Eqs. (\ref{eq:greens_definition}), using the anomalous Green's function $F_{nm}(\tau)=-\langle T_\tau c_{n}(\tau) c_{m}(0) \rangle$. It is straightforward to write down the equations of motion for these Green's functions as a generalized version of Eq. (\ref{eq:2band_eom}), from which we find that $\hat{F}$ is given by:
\begin{equation}
\begin{aligned}
\hat{F}(i\omega_n)= &\left[\left(i\omega_n-\hat{h}\right)-\hat{\Delta}\left(i\omega_n+\hat{h}^*\right)^{-1}\hat{\Delta}^\dagger \right]^{-1}\hat{\Delta} \\
&\times\left(i\omega_n+\hat{h}^*\right)^{-1}.
\end{aligned}
\label{eq:f_general}
\end{equation}
Here, the $\hat{}$ -symbol denotes matrices with indices $n$,$m$ running over all quantum numbers describing the quasiparticles, including position, spin, and any band or similar degrees of freedom. 

From Eq. (\ref{eq:f_general}) we see that, in general, this matrix should possess both even-$\omega$ and odd-$\omega$ terms, with the details depending on the precise form of the Hamiltonian. Further insight can be gained by expanding the right-hand-side to leading order in $\hat{\Delta}$, in which case we find linearized expressions for both the even- and odd-$\omega$ pair amplitudes:
\begin{equation}
\begin{aligned}
\hat{F}_{even}(i\omega_n)= &-\left[\omega_n^2+\hat{h}^2\right]^{-1}\left[\hat{h},\hat{\Delta}\right]_*\hat{h}^*\left[\omega_n^2+(\hat{h}^*)^2\right]^{-1}\\
&-\left[\omega_n^2+\hat{h}^2\right]^{-1}\hat{\Delta} , \\
\hat{F}_{odd}(i\omega_n)= &i\omega_n\left[\omega_n^2+\hat{h}^2\right]^{-1}\left[\hat{h},\hat{\Delta}\right]_*\left[\omega_n^2+(\hat{h}^*)^2\right]^{-1},
\end{aligned}
\label{eq:foddeven_general}
\end{equation}
where we define $\left[\hat{h},\hat{\Delta}\right]_*\equiv \hat{h}\hat{\Delta}-\hat{\Delta}\hat{h}^*$. 

Clearly, when $\left[\hat{h},\hat{\Delta}\right]_*$ vanishes, the system has only even-$\omega$ pairing, given by $-\left[\omega_n^2+\hat{h}^2\right]^{-1}\hat{\Delta}$. Thus the condition for the emergence of odd-$\omega$ pairing, in a general mean field theory, is given by:
\begin{equation}
\hat{h}\hat{\Delta}-\hat{\Delta}\hat{h}^* \neq 0.
\label{eq:odd_criterion}
\end{equation} 
For the simple two-band model in Eq. (\ref{eq:ham2band_compact_definitions}) this condition is clearly satisfied because $\hat{\Delta}$ and $\hat{h}$ are proportional to different $2\times 2$ Pauli matrices in band-space, but Eq.~\eqref{eq:odd_criterion} is much more general. In particular, the connection between the structures of $\hat{\Delta}$ and $\hat{h}$ and the emergence of odd-$\omega$ pairing applies to superconductors with any number of bands or other internal electronic degrees of freedom. For example, if $\hat{h}$ describes a generic real-space tight-binding Hamiltonian then, if $\hat{\Delta}$ is an inhomogeneous on-site order parameter, then the inequality in Eq. (\ref{eq:odd_criterion}) is generically satisfied and we expect to find odd-$\omega$ pairing. Note that this is entirely consistent with previous results studying the emergence of odd-$\omega$ pairing in inhomogeneous systems, including at SN interfaces \cite{tanaka2007theory, TanakaPRB2007,Black-SchafferPRB2012,cayao2017odd,triola2018oddnw,triola2019odd}. In contrast, if $\hat{\Delta}$ is a spatially homogeneous on-site order parameter and $\hat{h}$ is trivial in spin space and has only real elements, no odd-$\omega$ pairing is possible.  

If the system under consideration is translation invariant we can Fourier-transform from real-space to momentum space and we find the following condition for odd-$\omega$ pairing: 
\begin{equation}
\hat{h}_{\textbf{k}}\hat{\Delta}_\textbf{k}-\hat{\Delta}_\textbf{k}\hat{h}^*_{-\textbf{k}}\neq 0.
\label{eq:odd_criterion_k}
\end{equation}
This condition is in fact identical to a measure of ``superconducting fitness" recently discussed by Ramires and Sigrist\cite{ramires2016identifying}. In that work, it was demonstrated that whenever this quantity is non-zero there is a reduction in the critical temperature. The authors, therefore, concluded that superconducting fitness can be a tool in the search for order parameters that are expected to be more thermodynamically stable\cite{ramires2016identifying}. 
It is here interesting to note that, just one year prior to the publication of Ref. \cite{ramires2016identifying}, a work by Asano and Sasaki\cite{asano2015odd} concluded that the emergence of odd-$\omega$ pairing in two-band superconductors is linked to a suppression of the critical temperature. Given the general nature of the results by Ramires and Sigrist, together with the results presented in this section, we conclude that the assertions of Asano and Sasaki are likely to hold in general, i.e. the emergence of odd-$\omega$ pairing appears to cause a suppression of the superconducting critical temperature. 
It is here important to note that, even though the presence of odd-$\omega$ pairing is associated with a suppressed critical temperature, such a state can still easily be the most thermodynamically favored as that depends crucially on the form of the interaction and the normal state Hamiltonian. In the next section we discuss several examples of real systems believed to host exactly this kind of odd-$\omega$ pairing.

\section{Examples of Multiband Odd-frequency Pairing}
\label{sec:examples}

Having derived the general criteria for odd-$\omega$ superconductivity to appear in multiband systems, we now discuss real examples of multiband superconductors in which odd-$\omega$ pairing has been predicted to emerge. We begin by covering examples in which band degrees of freedom are intrinsic to the superconductor, arising from either different atomic orbitals or a sublattice index, forming what would properly be known as a multiband superconductor. We then discuss superconducting systems in which the additional electronic degrees of freedom are not strictly speaking band indices but have their origin in some other aspect of the system, considering the cases of two-dimensional (2D) bilayers, one-dimensional (1D) nanowires, zero-dimensional (0D) quantum dots, superconducting leads in Josephson junctions, and isolated valleys in momentum space.

\subsection{Sr$_{2}$RuO$_4$}
\label{sec:sr2ruo4}
In this subsection we discuss the emergence of odd-$\omega$ pairing in the multiband superconductor Sr$_{2}$RuO$_4$, which was recently examined by Komendov\'{a} and Black-Schaffer\cite{komendova2017odd}. While it possesses a fairly low critical temperature, $T_c\approx 1$K, the superconducting phase of Sr$_{2}$RuO$_4$ has attracted a great deal of attention since its discovery in 1994\cite{maeno1994superconductivity} due to its highly unusual properties. Both Knight shift and neutron scattering measurements have indicated the possibility of spin-triplet pairing \cite{ishida1998spin,ishida2015spin,duffy2000polarized}. Additionally, it has been observed that the superconducting phase exhibits spontaneous time-reversal symmetry breaking using muon spin-relaxation measurements\cite{luke1998time,luke2000unconventional}, as well as measurements of the Kerr effect\cite{xia_prl_2006}. Taken together, these heavily imply a chiral $p$-wave order parameter. However, measurements of the specific heat\cite{nishizaki1999effect,nishizaki2000changes,deguchi2004gap} are more consistent with a nodal gap structure. Furthermore, recent NMR studies revisiting the Knight shift have found evidence more consistent with spin-singlet pairing\cite{pustogow2019pronounced}. The lack of consistency between these complementary studies continues to make Sr$_{2}$RuO$_4$ both an interesting and hotly debated superconductor.

While the superconducting state of Sr$_{2}$RuO$_4$ is controversial, the normal state properties are now quite well-understood with experiments\cite{mackenzie1996quantum,bergemann2000detailed} and theory\cite{oguchi1995electronic,singh1995relationship} converging on the same picture of three quasi-2DFermi sheets, with contributions primarily from the ruthenium $d_{xy}$, $d_{xz}$, and $d_{yz}$ orbitals. Therefore, to capture the relevant physics of Sr$_{2}$RuO$_4$ a three-orbital Hamiltonian, similar to Eq. (\ref{eq:ham2band_compact}), can be employed, with normal state Hamiltonian, $\hat{h}$, and order parameter, $\hat{\Delta}$, given by:
\begin{equation}
\begin{aligned}
\hat{h}_{\textbf{k}}&=\left(\begin{array}{ccc}
\xi_1 & \epsilon_{12} & \epsilon_{12} \\
\epsilon_{12} & \xi_2 & \epsilon_{23} \\
\epsilon_{13} & \epsilon_{23} & \xi_3 
\end{array} \right), \ \hat{\Delta}_{\textbf{k}}&= \left(\begin{array}{ccc}
\Delta_1 & \Delta_{12} & \Delta_{13} \\
\Delta_{12} & \Delta_2 & \Delta_{23} \\
\Delta_{13} & \Delta_{23} & \Delta_{3}
\end{array} \right),
\end{aligned}
\label{eq:tbsr2ruo4}
\end{equation} 
where the $k$-dependence of the matrix elements has been suppressed for brevity and where the indices 1, 2, and 3 correspond to the ruthenium $d_{xy}$, $d_{xz}$, and $d_{yz}$ orbitals, respectively. Here, it is assumed that the order parameter is either spin-singlet or mixed spin-triplet\cite{komendova2017odd}.

Before assuming precise values for the tight-binding model in Eq. (\ref{eq:tbsr2ruo4}), two special cases were considered analytically in Ref.\cite{komendova2017odd}. For the first case, the order parameter was assumed to be completely diagonal in the orbital basis, $\hat{\Delta}=\text{diag}\left(\Delta_1,\Delta_2,\Delta_3\right)$, and the interorbital terms in $\hat{h}$ were all assumed to be equal, $\epsilon_{ij}=\Gamma$. The analytic expressions for the pair amplitudes were examined\cite{komendova2017odd} and it was found that the odd-$\omega$ pairing is present as long as at least two of the gaps are different, $\Delta_i\neq\Delta_j$ for some $i\neq j$. For the second case, the hybridization was assumed to only occur between the $d_{xz}$, and $d_{yz}$ orbitals, so that: $\epsilon_{12}=\epsilon_{13}=0$ and $\Delta_{12}=\Delta_{13}=0$, which is often used as as simplification. In this case, two separate contributions to the odd-$\omega$ pair amplitudes were found, one proportional to the interorbital component of the order parameter, $\sim \Delta_{23}(\xi_3-\xi_2)$, and one proportional to the interorbital hybridization, $\sim\epsilon_{23}(\Delta_3-\Delta_2)$. It is straightforward to confirm that these same conditions can be obtained from the criterion in Eq. (\ref{eq:odd_criterion_k}).

\begin{figure*}[htb]
\centering
\includegraphics*[width=0.8\textwidth]{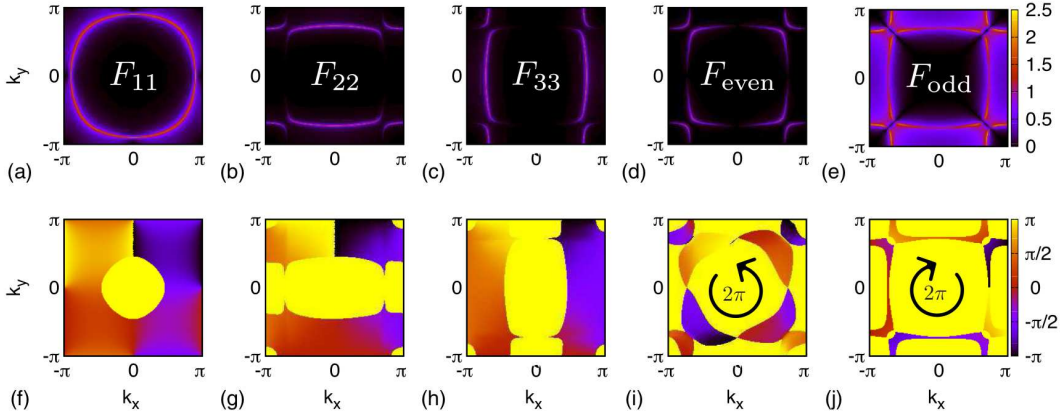}
\caption{Intraorbital pair amplitudes, $F_{11}$, $F_{22}$, $F_{33}$, and interorbital pair amplitudes, $F_{even}$ and $F_{odd}$, plotted over the first Brillouin zone for Sr$_2$RuO$_4$. Note that, $F_{11}$ possessses all of its spectral weight on the $\gamma$ band, while $F_{22}$, $F_{33}$, $F_{even}$ and $F_{odd}$, possess spectral weight on both the $\alpha$ and $\beta$ bands, consistent with the form of the interorbital hybridization. Top row represents the magnitudes of each function while bottom row shows the complex phase. For $F_{odd}$ the results are multiplied by a factor of 100. Reprinted figure with permission from [L. Komendov\'{a} and A. M. Black-Schaffer, Phys. Rev. Lett. 119, 087001 (2017)] Copyright (2017) by the American Physical Society.}
\label{fig:komendova_prl_2017_fig1}
\end{figure*}

Focusing specifically on parameters which faithfully reproduce the three bands of Sr$_2$RuO$_4$, $\gamma$, $\alpha$, and $\beta$, Ref.~\cite{komendova2017odd} also provided a numerical evaluation of all even- and odd-$\omega$ components of the anomalous Green's functions. In this calculation, the $\gamma$ band was assumed to have contributions only from orbital 1 ($d_{xy}$) while the $\alpha$ and $\beta$ bands emerge from hybridization between orbital 2 and orbital 3 ($d_{xz}$ and $d_{yz}$). Each of these channels was summed over the positive Matsubara frequencies and plotted over the first Brillouin zone, with the result presented in Fig. \ref{fig:komendova_prl_2017_fig1}, which is adapted from Ref.~\cite{komendova2017odd}. From these color plots we see that both the even- and odd-$\omega$ interorbital pair amplitudes, $F_{even}$ and $F_{odd}$, possess all of their weight along the same bands as $F_{22}$ and $F_{33}$, $\alpha$ and $\beta$. Additionally, the phases associated with $F_{even}$ and $F_{odd}$ undergo a full $2\pi$ rotation around the $\Gamma$ point, consistent with the assumed chiral $p$-wave order parameter. However, from the analytic criteria discussed in the previous paragraph, we see that assuming another order parameter will not change the results significantly. This confirms that, regardless of the precise symmetry of the order parameter in Sr$_2$RuO$_4$, it is likely to host odd-$\omega$ interband pairing due to interband hybridization.  

\subsection{UPt$_3$}
\label{sec:upt3}
Next we discuss a recent work demonstrating the emergence of odd-$\omega$ pairing in the heavy-fermion superconductor UPt$_3$\cite{triola2018odd}. In addition to possessing multiple relevant bands at the Fermi level, UPt$_3$ is a truly unconventional superconductor, exhibiting two zero-field superconducting phases, the $A$ phase and the $B$ phase, with critical temperatures $T_{c,+}\approx 550$ mK and $T_{c,-}\approx 500$ mK\cite{stewart_prl_1984,sauls1994}, respectively. Additionally, a third phase, the $C$ phase, emerges at high magnetic field\cite{adenwalla1990}. Knight shift observations point to a spin-triplet superconducting order parameter\cite{tou_prl_1996}. Josephson interferometry has revealed the presence of line nodes in the A phase\cite{strand_science_2010}, as well as the onset of a complex order parameter in the $B$ phase\cite{strand_prl_2009,strand_science_2010}. Moreover, recent measurements of the Kerr effect have demonstrated time-reversal symmetry breaking in the $B$ phase, consistent with a complex order parameter \cite{schemm_2014}. 

To capture the essential features of the Fermi surfaces appearing at either the $\Gamma$-point or the $A$-point in UPt$_3$ the following normal state tight-binding Hamiltonian has recently been employed\cite{yanase_prb_2016, yanase_2017_prb,wang_2017}:
\begin{equation}
\begin{aligned}
\hat{h}_{\textbf{k}}&=\left( \begin{array}{cccc}
\xi_\textbf{k} + g_\textbf{k} & \epsilon_\textbf{k} & 0 & 0 \\
\epsilon_\textbf{k}^* & \xi_\textbf{k} - g_\textbf{k} & 0 & 0 \\
0 & 0 & \xi_\textbf{k} - g_\textbf{k} & \epsilon_\textbf{k} \\
0 & 0 & \epsilon_\textbf{k}^* & \xi_\textbf{k} + g_\textbf{k}
\end{array}\right), 
\end{aligned}
\label{eq:hamiltonian}
\end{equation}
written in the basis described by $\Psi^\dagger=( c^\dagger_{\textbf{k}1\uparrow}, c^\dagger_{\textbf{k}2\uparrow}, c^\dagger_{\textbf{k}1\downarrow}, c^\dagger_{\textbf{k}2\downarrow})$ where $c^\dagger_{\textbf{k}m\sigma}$ creates a fermionic quasiparticle with crystal momentum $\textbf{k}$, on sublattice $m=\{1,2\}$, and with spin $\sigma=\{\uparrow,\downarrow\}$. Here, $\xi_\textbf{k}$ is an even function of $\textbf{k}$ describing the intra-sublattice hopping, $\epsilon_\textbf{k}$ is a complex-valued inter-sublattice hopping term, and the function $g_\textbf{k}$ is odd in $\textbf{k}$ and describes the spin-orbit coupling. In Eq. (\ref{eq:hamiltonian}) we note that, in contrast to Sr$_2$RuO$_4$ whose multiband character has its origin in the atomic orbitals of the Ru atoms, the multiple bands within this model have their origin in the sublattice degree of freedom with contributions coming from only a single itinerant $5f$ orbital of the uranium atoms. 

The superconducting order parameter in UPt$_3$ is widely believed to belong to the $E_{2u}$ irreducible representation with spin-triplet $m_z = 0$ pairing\cite{sauls1994,joynt_rmp_2002,yanase_prb_2016}. Following recent work explicitly accounting for the symmetries of the lattice \cite{yanase_prb_2016}, the order parameter is given by a linear combination of $d$-wave and $f$-wave basis functions:
\begin{equation}
\hat{\Delta}_{\textbf{k}}= f_\textbf{k} \hat{\rho}_1\otimes\hat{\sigma}_1 - d_{\textbf{k}}\hat{\rho}_2\otimes\hat{\sigma}_1,
\label{eq:upt3gap}
\end{equation}
where $\hat{\sigma}_i$ and $\hat{\rho}_i$ are Pauli matrices in spin and sublattice space, respectively, $f_\textbf{k}=\eta_1 f_{(x^2-y^2)z}(\textbf{k})+\eta_2f_{xyz}(\textbf{k})$ and $d_\textbf{k}=\eta_1 d_{yz}(\textbf{k})+\eta_2d_{xz}(\textbf{k})$, and $\eta_i$ are complex numbers parameterizing the phase diagram\cite{sauls1994,joynt_rmp_2002,nomoto_prl_2016,yanase_prb_2016}.
Notice the unusual combination of spin-triplet $f$-wave terms being odd in spatial parity and spin-triplet $d$-wave terms being even in parity. This combination is caused by the nonsymmorphic lattice symmetry \cite{yanase_prb_2016}. Note that these terms still satisfy the constraints imposed by Fermi-Dirac statistics on the Cooper pairs since the $f$-wave terms are even in the sublattice index while the $d$-wave terms are odd in the sublattice index, belonging to the symmetry classes in columns 2 and 3 of Table \ref{table:classification}, respectively, when viewing the sublattice index as a band degree of freedom. 

Using the Hamiltonian and order parameter in Eqs. (\ref{eq:hamiltonian}) and (\ref{eq:upt3gap}), the symmetries of the anomalous Green's function were explored in Ref.~\cite{triola2018odd}, using the same conventions as in Sec.~\ref{sec:gen}. In that analysis UPt$_3$ was found to exhibit a plethora of pairing channels, both even and odd in frequency. More specifically, four different kinds of odd-$\omega$ pair amplitudes were found, with the general form\cite{triola2018odd}:
\begin{equation}
\begin{aligned}
\hat{F}_{odd}&=\psi_1 \hat{\rho}_3\otimes\hat{\sigma}_1 + \psi_2\hat{\rho}_0\otimes\hat{\sigma}_2+ \psi_3\hat{\rho}_1\otimes\hat{\sigma}_2 + \psi_4\hat{\rho}_2\otimes\hat{\sigma}_2.
\end{aligned}
\label{eq:fodd_upt3}
\end{equation} 
From the matrix structure in Eq. (\ref{eq:fodd_upt3}) we see that the intra-sublattice spin-triplet amplitude, $\psi_1$, corresponds to the symmetry class in column 5 of Table \ref{table:classification}, while the intra-sublattice spin-singlet amplitude, $\psi_2$, and the even inter-sublattice spin-singlet amplitude, $\psi_3$, both correspond to the symmetry class in column 6. Finally, the odd inter-sublattice spin-singlet amplitude, $\psi_4$, belongs to the symmetry class in column 7. In terms of the physical parameters, the presence of a finite inter-sublattice term, $\epsilon_\textbf{k}\neq 0$, gives rise to the odd-$\omega$ intra-sublattice term $\psi_1$, despite the fact that the initial order parameter in Eq. (\ref{eq:upt3gap}) is entirely in the inter-sublattice channels. Moreover, the addition of spin-orbit coupling, $g_\textbf{k}$, gives rise to multiple odd-$\omega$ spin-singlet inter-sublattice pair amplitudes, one of which is sublattice odd, $\psi_3$, and the other sublattice even, $\psi_4$. Finally, the combination of both spin-orbit coupling and inter-sublattice hybridization leads to the odd-$\omega$ intra-sublattice spin-singlet term, $\psi_2$.

\subsection{Buckled Honeycomb Materials}
In this subsection we discuss the emergence of odd-$\omega$ pairing in buckled 2D honeycomb lattices with proximity-induced superconductivity, investigated in Refs.~\cite{black2013odd, kuzmanovski2017multiple}. It is well-known that 2D honeycomb lattices are composed of two triangular sublattices\cite{katsnelson2012graphene}. This sublattice degree of freedom gives rise to two bands near the Fermi level, similar in spirit to the multiband nature of UPt$_3$ discussed in the previous subsection. But more remarkable in honeycomb materials is that the intersublattice hybridization is especially prominent, since the dominating nearest neighbor hopping necessarily couple the two sublattices. This band structure is realized in many of the known 2D materials, including graphene\cite{novoselov2004electric,neto2009electronic,katsnelson2012graphene}, silicene\cite{vogt2012silicene}, germanene\cite{liu2011quantum,davila2014germanene}, and stanene\cite{zhu2015epitaxial}. While the two sublattices in graphene are symmetric and lie in the same plane, in silicene, germanene, and stanene, the structures are naturally buckled, so that the two sublattices are staggered. Therefore, in the latter three materials an asymmetry between the two sublattices can be induced and controlled simply by applying a gate voltage perpendicular to the layer. Such an asymmetry between the sublattices has been shown to directly lead to odd-$\omega$ pairing in these materials \cite{black2013odd}, in complete analogy with the results in Sec.~\ref{sec:twoband_criteria}. Another interesting aspect of buckled honeycomb materials is that a sublattice asymmetry has also been shown to appear in finite-width nanoribbons due to the presence of sample edges \cite{kuzmanovski2017multiple}.    

More specifically, in Ref.~\cite{kuzmanovski2017multiple}, the authors start by describing the normal state of a buckled honeycomb system with possibly finite spin-orbit coupling, using the Kane-Mele Hamiltonian in real space\cite{haldane1988model,kane2005qshi}:
\begin{equation}
\begin{aligned}
H_0&=t\sum_{\langle i,j\rangle,\sigma} c^\dagger_{i\sigma}c_{j\sigma}+\frac{i\lambda_{\text{SO}}}{3\sqrt{3}}\sum_{\langle\langle i,j\rangle\rangle,\sigma}\nu_{ij}(\hat{\sigma}_3)_{\sigma\sigma'} c^\dagger_{i\sigma}c_{j\sigma'} \\
&-\sum_{i,\sigma}\mu_i c^\dagger_{i\sigma}c_{i\sigma},
\end{aligned}
\label{eq:kane-mele}
\end{equation} 
where $c^\dagger_{i\sigma}$ ($c_{i\sigma}$) creates (annihilates) as fermionic quasiparticle at site $i$ with spin $\sigma$, $\langle i,j\rangle$ sums over nearest-neighbor (NN) sites, $i,j$, of the honeycomb lattice, $\langle\langle i,j\rangle\rangle$ sums over next-nearest-neighbor (NNN) sites. Here, $t$ represents the NN hopping parameter and $\lambda_{\text{SO}}$ is the spin-orbit coupling due to NNN hopping, where $\nu_{ij}=\pm 1$ depending on whether the vector from site $i$ to $j$ is oriented clockwise or counterclockwise around the hexagonal plaquette\cite{haldane1988model}. The possibility of gating, is captured by a sublattice-dependent chemical potential $\mu_i=\mu+\zeta_i \lambda_V$, where $\mu$ is the chemical potential in the absence of any applied voltage, $\lambda_V$ is proportional to the applied voltage, and $\zeta_i=\pm 1$ depending on whether $i$ belongs to sublattice A or B.  

At finite doping, the normal state described by Eq. (\ref{eq:kane-mele}) possesses a large enough electronic density of states for bulk superconductivity to be induced by proximity effect. In this limit, the bulk pair amplitudes have been studied by transforming the above model to momentum space and assuming a $k$-independent $s$-wave order parameter, as appropriate for proximity effect from a conventional superconductor \cite{kuzmanovski2017multiple}. The resulting Hamiltonian possesses a similar form to the two-band model in Eq. (\ref{eq:ham2band_compact}) but with a momentum-dependent interband hybridization term and non-trivial spin structure parameterized by $\lambda_{\text{SO}}$. Solving for the anomalous Green's function, odd-$\omega$ pairing in all four odd-$\omega$ symmetry classes in Table~\ref{table:classification} are possible in this system, although the authors of Ref. \cite{kuzmanovski2017multiple} did not mention the pair amplitudes belonging to columns 5 and 6 in their discussion. In particular, odd-$\omega$ spin-singlet pair amplitudes are present whenever there is an asymmetry between the order parameters on the different sublattices, i.e. $\Delta_{\text{A}}\neq\Delta_{\text{B}}$, with both even- and odd-sublattice contributions due to the momentum-dependent intersublattice NN hopping. The required order parameter sublattice difference is present when $\lambda_V\neq 0$ and thus odd-$\omega$ pairing is controlled by gating\cite{kuzmanovski2017multiple}. 
Moreover, odd-$\omega$ spin-triplet pairing is present due to a finite spin-orbit coupling, $\lambda_\text{SO}\neq 0$.

As is well-known, for $\lambda_V<\lambda_\text{SO}$ the Kane-Mele Hamiltonian in Eq.~\eqref{eq:kane-mele} describes a topological insulator with a bulk band gap and conducting edge modes. Ref.\cite{kuzmanovski2017multiple} also studied this phase by considering nanoribbons with both zigzag (ZZ) and armchair (AC) terminations in the low doping reigme. In this case, superconductivity vanishes throughout the bulk, but a finite $\Delta_i$ was obtained using a self-consistent algorithm for each site along the edges\cite{kuzmanovski2017multiple}. 
However, in contrast to the translation-invariant case, the magnitudes of all pair amplitudes in these cases are largest in the absence of $\lambda_V$. Still, odd-$\omega$ pairing appear in these ribbons due to an inherent asymmetry between the two sublattices at the edges. In the case of the ZZ termination, the A and B sublattices are clearly different at the edge, since one sublattice has only two NNs while the other retains three. For AC termination, the situation is a bit more subtle as the two sublattices are equivalent, but an asymmetry exists between every other pair of sublattices. The latter induces a gradient of the order parameter along the edge, which is also known to induce odd-$\omega$ pairing in topological insulators \cite{Black-SchafferPRB2012}. 

\subsection{Other Analogous Systems}
\label{sec:other}
One common aspect of the previously discussed examples is that, odd-$\omega$ pairing emerges from the hybridization of a discrete set of multiple bands. These multiple bands offer an expansion of the set of allowed Cooper pair symmetries, as illustrated in Table \ref{table:classification}, and have their origin in either the atomic orbitals associated with individual lattice sites of a bulk crystal or the sublattice structure defining the crystal's unit cell. However, there are other ways to obtain similar discrete sets of multiple ``bands" in superconducting systems, as we now discuss. 

One proposal by Parhizgar and Black-Schaffer\cite{parhizgar_2014_prb} involves the use of 2D bilayer systems proximity-coupled to conventional superconductors, In this case the layer index provides a band-like degree of freedom analogous to the preceding examples. Such 2D bilayer systems include, bilayer graphene\cite{ohta2006controlling,sarma2011electronic,katsnelson2012graphene,mccann2013electronic}, bilayer transition metal dichalcogenides\cite{ramasubramaniam2011tunable,zhang2016visualizing}, other layered Van der Waals heterostructures\cite{geim2013van,novoselov20162d}, as well as topological insulator thin films\cite{zhang2010crossover,cheng2010landau,zhang2011growth}. These kinds of layered systems have uniquely tunable electronic properties due to the variety of 2D systems available, as well as the ability to control their electronic properties through gating and introducing a relative twist angle between the layers\cite{li2010observation,bistritzer2010transport,dos2012continuum,cao2018unconventional}. As shown in Ref. \cite{parhizgar_2014_prb} when a generic bilayer 2D system is proximity coupled to a conventional $s$-wave spin-singlet superconducting substrate, the layer closest to the substrate necessarily obtains a larger superconducting gap, thus directly producing a layer asymmetry. Further, when examining the symmetries of the anomalous Green's function, a rich variety of allowed symmetries were found, including both even- and odd-$\omega$ interlayer pairing. Moreover it was determined that within these models there is a complete reciprocity between the layer symmetry and the frequency symmetry: all odd-layer amplitudes are odd-$\omega$, all even-layer amplitudes are even-$\omega$\cite{parhizgar_2014_prb}, in complete analogy with results for two-band superconductors\cite{black2013odd}.

Another set of proposals rely on double-quantum dots coupled to superconductors\cite{sothmann2014unconventional,burset2016all}. In this case the dot index acts as an effective band index and interdot coupling can thus induce odd-$\omega$ pairing. The first proposal by Sothmann and collaborators\cite{sothmann2014unconventional} utilized two quantum dots proximitized by a conventional $s$-wave superconductor, in the presence of both interdot tunneling and an external magnetic field. They demonstrated a variety of possible odd-$\omega$ pair amplitudes in these systems, both spin-singlet and spin-triplet, and tunable using either the externally applied magnetic field, a difference in on-site energy levels, or an asymmetry in coupling between normal and superconducting leads\cite{sothmann2014unconventional}. This possibility was explored further by Burset and colleagues\cite{burset2016all} in the absence of a magnetic field. In this case, they were able to find spin-triplet pairing by coupling the two dots to a spin-triplet superconductor. Both studies also explored tunable signatures of the odd-$\omega$ pairing observable in transport between superconducting or normal leads\cite{sothmann2014unconventional,burset2016all}.  

In a similar spirit to the proposals involving double quantum dots, odd-$\omega$ pairing has also been proposed in double nanowires coupled to a superconducting substrate\cite{ebisu2016theory,triola2018oddnw}. In these cases, the nanowire index acts as an effective band. In a work by Ebisu {\it et al.} \cite{ebisu2016theory}, an effective model was used to describe two nanowires with Rashba spin-orbit coupling in the presence of both intrawire and interwire superconducting mean fields. It was found that, for generic parameters, odd-$\omega$ pairing is present in these systems, and that it is strongly enhanced when the system is tuned into the topological regime, where interwire pairing dominates. In a later work, Triola and Black-Schaffer \cite{triola2018oddnw} studied a similar setup but explicitly considered the two nanowires coupled to a 2D superconductor and studied the emergent pair amplitudes of this system as a whole. In particular, they found that, in agreement with previous work, odd-$\omega$ interwire pairing is generically induced by coupling the two wires to the superconductor. Moreover, the authors showed that the presence of the nanowires also profoundly affect the pair symmetries of the superconducting substrate, leading to measurable signatures in local observables\cite{triola2018oddnw}.   

Odd-$\omega$ pairing has also been explored in conventional Josephson junctions\cite{linder2017odd,balatsky2018odd}, in which the two weakly-coupled superconducting leads naturally provides a lead index, playing the role of bands. Interestingly, it was found that, in general, Josephson junctions should possess odd-$\omega$ interlead pairing proportional to $\sin{\tfrac{\phi}{2}}$, where $\phi$ is the phase difference across the junction. Comparing this condition to the well-known formula for the Josephson current, it was concluded that whenever Josephson current is expected to flow across the junction odd-$\omega$ interlead pairing will also be present.

The above examples involving bilayers, double quantum dots, double nanowires, and Josephson junctions, all utilize spatial separation to obtain an additional index akin to the band index, but it is also possible to obtain such an index using a separation in reciprocal space. In particular, the transition metal dichalcogenides (TMDs) may be described by an effective model governing the physics of separate points in the Brillouin zone, so-called valleys. In these systems, the valley index can thus behave like an effective band degree. Using a low-energy effective model to describe the two $k$-space valleys of a single layer of TMD proximity-coupled to an $s$-wave superconductor with Rashba spin-orbit coupling, Ref. \cite{triola2016prl} found that the combination of valley-dependent spin-orbit coupling, intrinsic to the monolayer TMD, and the Rashba spin-orbit term at the TMD-superconductor interface necessarily leads to an odd-$\omega$ intervalley pair amplitude.

\section{Experimental Signatures}
\label{sec:experiment}
Having shown how odd-$\omega$ superconductivity is ubiquitous in many superconducting systems, we now present several experimental signatures that have been proposed to measure the odd-$\omega$ pairing. Due to its intrinsically dynamical nature, with a zero equal-time amplitude, odd-$\omega$ pairing has proven to be notoriously hard to probe directly, still, as seen below, there are a growing number of known signatures of odd-$\omega$ pairing in multiband superconductors.

\subsection{Hybridization Gaps}
Shortly after the initial theoretical proposal for the emergence of odd-$\omega$ pairing in the two-band model defined in Eq. (\ref{eq:ham2band_compact})\cite{black2013odd} it was observed that the emergence of interband odd-$\omega$ pairing can be correlated with measurable signatures in the density of states (DOS)\cite{komendova2015experimentally}. In Ref. \cite{komendova2015experimentally} the simple two-band Hamiltonian, Eq. (\ref{eq:ham2band_compact}), was considered and the DOS was computed to search for features correlated with the emergence of odd-$\omega$ pairing. In addition to the total DOS, the separate contributions to the DOS coming from bands 1 and 2, $N_1$ and $N_2$, were examined to highlight the features which are strictly intraband and those which obtain contributions from both.

\begin{figure}
 \begin{center}
  \centering
  \includegraphics[width=0.5\textwidth]{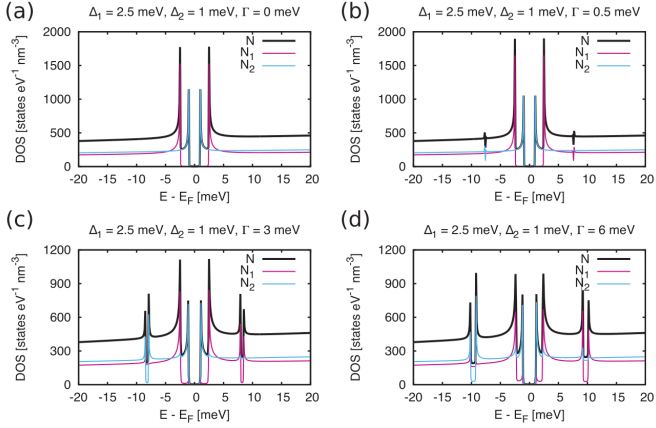}
  \caption{DOS computed for the two-band model in Eq. (\ref{eq:ham2band_compact}) using $m_1=20m_e$, $m_2=22m_e$, $\mu_1=100 \text{meV}$, $\mu_2=105 \text{meV}$, $\Delta_1=2.5 \text{meV}$, and $\Delta_2=1 \text{meV}$, for four values of $\Gamma$, specified in each panel. Reprinted figure with permission from [L. Komendov\'{a}, A. V. Balatsky, and A. M. Black-Schaffer, Phys. Rev. B 92, 094517 (2015)] Copyright (2015) by the American Physical Society.}
  \label{fig:ldos_2band}
 \end{center}
\end{figure}

Computing the DOS in this manner, it was found that, as expected, in the presence of two different gaps $\Delta_1\neq\Delta_2$, and in the absence of interband hybridization, $\Gamma=0$, the DOS is simply a superposition of the DOS of two superconductor with coherence peaks at $E=\pm\Delta_1$ and $E=\pm\Delta_2$, respectively, see Fig. \ref{fig:ldos_2band}(a). However, for any finite value of $\Gamma$ additional gaps, and associated coherence peaks, were found to appear at energies away from the two gaps at the Fermi level, Figs. \ref{fig:ldos_2band}(b)-(d). 
These hybridization-induced gaps arise due to avoided crossings in the quasiparticle dispersion at the energies where the bands $E_1=\sqrt{\xi_1^2+|\Delta_1|^2}$ and $E_2=\sqrt{\xi_2^2+|\Delta_2|^2}$ meet. Solving for these crossing points, it can be shown that, in general, there could be two avoided crossings at different positive energies. However, assuming the initial quasiparticle bands do not intersect, i.e. $\xi_{1,\textbf{k}}\neq \xi_{2,\textbf{k}}$ for any $\textbf{k}$, then only one of these is a true avoided crossing. For quadratic dispersions with effective masses $m_i$ and chemical potentials $\mu_i$, this condition will be true as long as $(\mu_1-\mu_2)/(m_1-m_2)>0$. In this case, one can show that hybridization gaps will emerge if and only if $\Gamma\neq 0$ and $\Delta_1\neq\Delta_2$, which are exactly the same as the conditions for odd-$\omega$ interband pairing \cite{komendova2015experimentally}. 

While, hybridization gaps are a robust and simple probe of odd-$\omega$ pairing in the kind of two band model considered in Ref. \cite{komendova2015experimentally}, they do not emerge in all models for multiband superconductors with odd-$\omega$ pairing, since their emergence requires that the Bogoliubov bands intersect, in the absence of interband hybridization. For example, in the case of UPt$_3$, the hybridization term, $\epsilon_\textbf{k}$, induces odd-$\omega$ pair amplitudes; however, from the Hamiltonian in Eq. (\ref{eq:hamiltonian}) no avoided crossings emerge due to $\epsilon_\textbf{k}$, because, in the absence of spin-orbit coupling ($g=0$) the bands are degenerate for $\epsilon_\textbf{k}=0$ \cite{triola2018odd}. Furthermore, it can also be shown that neither Sr$_2$RuO$_4$\cite{komendova2017odd} nor the buckled honeycomb lattice\cite{kuzmanovski2017multiple} possess these hybridization gaps for similar reasons. Thus, it is necessary to also study alternative experimental signatures of odd-$\omega$ pairing in multiband superconductors.        

\subsection{Paramagnetic Meissner Effect}
\label{sec:paramagnetic}
One of the defining properties of superconducting states is their response to magnetic fields. As was first discovered by Meissner and Ochsenfeld\cite{meissner1933neuer}, superconductors exhibit perfect diamagnetism, referred to as the Meissner effect, in which magnetic flux is completely expelled from the bulk of a superconductor\cite{tinkham2004introduction,abrikosov2012methods}. In contrast to these classic results, it has been established by numerous theoretical works that odd-$\omega$ pairing often attracts magnetic flux, in a phenomenon termed the \textit{paramagnetic} Meissner effect \cite{tanaka2005anomalous, asano2011unconventional, higashitani2013magnetic, asano2014consequences, asano2015odd, FominovPRB2015} to contrast with the usual \textit{diamagnetic} Meissner effect. Such a paramagnetic response has been observed experimentally in magnetic-superconductor junctions using $\mu$SR, demonstrating that long-lived odd-$\omega$ pair amplitudes dominate deep within the magnetic bulk\cite{di2015intrinsic}. 

In Ref. \cite{asano2015odd} the magnetic response was studied using a two-band model similar to the one in Eq. (\ref{eq:ham2band_compact_definitions}), but with normal Hamiltonian possessing two kinds of interband hybridization, one spin-independent hybridization similar to $\Gamma$, and a spin-dependent hybridization with components given by $-\textbf{L}\times\textbf{k}\cdot\boldsymbol{\sigma}$, where $\textbf{L}$ describes the spin-orbit coupling in the system. The authors also considered three different types of order parameters: (i) spin-singlet even-parity intraband, (ii) spin-singlet even-parity even-interband, and (iii) spin-triplet even-parity odd-interband order. In each of these three cases it was found that odd-$\omega$ pairing can be induced by some asymmetry between the two bands, but the particular asymmetry and the properties of the induced odd-$\omega$ pairing were found to be different in each of case\cite{asano2015odd}. 

For the model in Ref. \cite{asano2015odd} the current density, $\textbf{j}$, can be related to a uniform applied magnetic field, $\textbf{A}$, with linear response theory: 
\begin{equation}
\textbf{j}=-K \textbf{A}
\end{equation} 
where $K$ is the Meissner kernel, which can be written in terms of the Nambu-Gorkov Green's functions: $\hat{G}$, $\hat{F}$, and $\hat{\bar{F}}$. Furthermore, assuming equal masses $m_1=m_2=m$ and chemical potentials $\mu_1=\mu_2=\mu$ for the two bands, the contribution to the Meissner kernel, $K$, takes on a relatively simple form:
\begin{equation}
K_F=\frac{e^2}{c} \frac{1}{m^2} T\sum_{\omega_n} \frac{1}{V_{\text{vol}}}\sum_\textbf{k}\frac{k^2}{d} \text{Tr}[\hat{F}(\textbf{k};i\omega_n)\hat{\bar{F}}(\textbf{k};i\omega_n)],
\label{eq:K_F}
\end{equation} 
where $e$ is the charge of the electron, $c$ the speed of light, $T$ the temperature, and $V_{\text{vol}}$ is the volume of the system in $d$ dimensions. 

Using Eq. (\ref{eq:K_F}), the authors examined the contributions to the Meissner effect coming from of each of the different superconducting pair channels. For case (i), and focusing on the simple case of $\textbf{L}=0$ and $\Gamma\neq 0$, only even-parity spin singlet pairing can emerge. Here, since only spin-singlet and even-parity pairing are induced, we find that $\hat{\bar{F}}(\textbf{k};i\omega_n)=-\hat{F}(\textbf{k};i\omega_n)^*$ and:
\begin{equation}
\begin{aligned}
\hat{F}(\textbf{k};i\omega_n)&=i\hat{\sigma}_2\otimes \sum_{i=0}^3 f_{i}(\textbf{k};i\omega_n) \hat{\rho}_i, \\
\end{aligned}
\label{eq:f_asano}
\end{equation}
where the odd-$\omega$ pair amplitude is necessarily given by the coefficient proportional to $\hat{\rho}_2$ (second Pauli matrix in band space), since that is the only possibility consistent with the symmetry constraints given by Fermi-Dirac statistics. From Eqs. (\ref{eq:K_F}) and (\ref{eq:f_asano}) it is easy to see that
\begin{equation}
\begin{aligned}
K_F=&\frac{e^2}{c} \frac{1}{m^2} T\sum_{\omega_n} \frac{1}{V_{\text{vol}}}\sum_\textbf{k}\frac{k^2}{d} \left[|f_0(\textbf{k};i\omega_n)|^2 \right. \\
&\left.+|f_1(\textbf{k};i\omega_n)|^2-|f_2(\textbf{k};i\omega_n)|^2 +|f_3(\textbf{k};i\omega_n)|^2 \right],
\end{aligned}
\end{equation} 
where all of the terms are strictly positive except for the contribution from the odd-$\omega$ pairing. This explicitly demonstrates that, in this case, odd-$\omega$ pairing always contributes paramagnetically to the Meissner kernel, thus countering the flux repulsion due to the conventional even-$\omega$ Cooper pairs. The authors went on to show that this pattern holds for all of the even-$\omega$ and odd-$\omega$ pair amplitudes in the three of the cases described above, demonstrating that, in a generic two-band model, all even-$\omega$ Cooper pairs exhibit diamagnetism while all odd-$\omega$ pairs exhibit paramagnetism\cite{asano2015odd}. 
Their analysis thus establishes the paramagnetic Meissner effect as a direct probe of odd-$\omega$ pairing in multiband systems. However, since both even- and odd-$\omega$ pair amplitudes are usually present and only the total Meissner response can be measured, isolation of the paramagnetic contributions may be challenging.

\subsection{Kerr Effect}
It has long been known that when polarized light is reflected from the surface of a magnetic material, the polarization of the reflected light can be shifted by an angle $\theta_{\text{K}}$ relative to the incident beam. This phenomenon, known as the Kerr effect, gives a direct probe of the breaking of time-reversal symmetry in magnetic materials. In recent years, the Kerr effect has also been applied to study time-reversal symmetry breaking (TRSB) order parameters in superconductors, in the absence of magnetism\cite{xia_prl_2006,schemm_2014}. However, it was later  established that multiband mechanisms are also necessary to observe the Kerr effect in clean superconductors even if the order parameter breaks TRSB \cite{taylor_prl_2012,taylor2013anomalous,wang_2017}. When applied to realistic tight-binding models, these calculations appear to match observations of the Kerr effect in both Sr$_2$RuO$_4$\cite{wysokinski_2012_prl,taylor_prl_2012,taylor2013anomalous,gradhand_2013_prb} and UPt$_3$\cite{wang_2017}. 

In particular, Taylor and Kallin\cite{taylor_prl_2012} studied the Kerr angle using a two-band model to describe superconducting Sr$_2$RuO$_4$. This model has the exact same form as Eq. (\ref{eq:ham2band_compact}) but with a real-valued momentum-dependent interband hybridization, $\Gamma_\textbf{k}$, and an order parameter that has both intraband components, $\Delta_{1}$ and $\Delta_2$, and an interband component $\Delta_{12}$. Using this model, they demonstrated that a necessary condition for the observation of a finite Kerr angle is:
\begin{equation}
\Gamma\text{Im}(\Delta_1^*\Delta_2)+\xi_{1}\text{Im}(\Delta_2^*\Delta_{12})-\xi_{2}\text{Im}(\Delta_1^*\Delta_{12})\neq 0,
\label{eq:2band_kerr}
\end{equation} 
where we have suppressed the $k$-dependence on the left-hand side for brevity. This implies that, in addition to a TRSB order parameter, either interband hybridization or a complex interband order parameter are essential for the observation of the Kerr effect in a clean two-band superconductor without magnetism. 

In Ref.\cite{komendova2017odd}, the criterion for a finite Kerr effect, Eq. (\ref{eq:2band_kerr}), was compared to the conditions for odd-$\omega$ pairing in that same model. There, it was demonstrated that whenever there is a finite Kerr effect, there will be odd-$\omega$ pairing in the system. The only possible exception was for the case in which $\xi_1=\xi_2$ and $\Delta_1\neq\Delta_2$; however, this would be incredibly unlikely. The same conclusion, that a finite Kerr effect signals the existence of odd-$\omega$ pairing, was also found to hold for a more realistic three-band model of Sr$_2$RuO$_4$ \cite{wysokinski_2012_prl,gradhand_2013_prb}. 
These results were later extended to UPt$_3$\cite{wang_2017,triola2018odd} demonstrating that the conditions giving rise to the Kerr effect are generically accompanied by odd-$\omega$ pairing. Taken together, these results solidify the status of the Kerr effect as a probe of odd-$\omega$ pairing in multiband superconductors with TRSB order parameters and strongly support the premise that both Sr$_2$RuO$_4$ and UPt$_3$, host odd-$\omega$ pairing. 
It is worth noting, however, that while these results show that the Kerr effect measures odd-$\omega$ pairing, it is possible to have odd-$\omega$ paring without exhibiting a Kerr effect, since the Kerr effect requires TRSB, which is only present in a few odd-$\omega$ multiband superconductors. Therefore, the lack of a finite Kerr angle is not evidence for the absence of odd-$\omega$ pairing.

\section{Conclusions}
\label{sec:conclusions}
In this article we have reviewed recent work on the possibility of odd-$\omega$ pairing in multiband superconductors. After a brief pedagogical examination of the emergence of odd-$\omega$ pairing in a simple two-band model we extended the formalism to derive a general criterion for the emergence of odd-$\omega$ pairing in any superconductor with an equal-time BCS order parameter, $\Delta$, and normal state Hamiltonian, $h$, given in Eq. (\ref{eq:odd_criterion}): $h\Delta-\Delta h^*\neq 0$. We noted that this condition is identical to a recently proposed measure of superconducting fitness which has been shown to suppress the superconducting critical temperature\cite{ramires2016identifying}. 

We then discussed several previous works in which multiband superconductors are predicted to host odd-$\omega$ pairing. In particular, we focused on Sr$_2$RuO$_4$\cite{komendova2017odd}, UPt$_3$\cite{triola2018odd}, and buckled honeycomb lattices\cite{kuzmanovski2017multiple}. In addition to these examples we also discussed several similar systems which have been predicted to host odd-$\omega$ pairing due to a band-like degree of freedom. These systems included proximitized bilayers\cite{parhizgar_2014_prb}, double quantum dots\cite{sothmann2014unconventional,burset2016all}, double nanowires\cite{ebisu2016theory,triola2018oddnw}, Josephson junctions\cite{linder2017odd,balatsky2018odd}, and monolayer transition metal dichalcogenides\cite{triola2016prl}. 

After discussing examples of systems which are predicted to host odd-$\omega$ pairing, we reviewed three different experimental probes which are relevant for odd-$\omega$ pairing in multiband systems: hybridization-induced gaps in the electronic density of states\cite{komendova2015experimentally}; paramagnetic Meissner effect\cite{asano2015odd}; and Kerr effect\cite{komendova2017odd,triola2018odd}. Each observable was found to have both distinct advantages and disadvantages. Hybridization-induced gaps always accompany odd-$\omega$ interband pairing in certain two-band models, thus providing a robust signature of odd-$\omega$ pairing. However, these gaps only appear when the Bogoliubov band structure exhibits specific avoided crossings and are therefore not observable in all multiband superconductors. A paramagnetic Meissner signal is a robust signature of odd-$\omega$ pairing, as it does not depend sensitively on the band structure. However, since even-$\omega$ pairing is expected to coexist with the odd-$\omega$ amplitudes, the net magnetic response is likely to be diamagnetic in generic multiband superconductors. Finally, a finite Kerr effect always signals odd-$\omega$ pairing, but only exists in superconductors which break time-reversal symmetry, which is not true for all odd-$\omega$ states.

To conclude, the ubiquity of odd-$\omega$ superconductivity has been shown in a wide variety of superconducting materials and systems, ranging from traditional multiband superconductors, to systems where other electronic degrees fo freedom provide an effective band index, including systems with layer, dot, wire, lead, and valley indices. Most importantly, as we demonstrated, the basic principles leading to the emergence of odd-$\omega$ pairing in all of these diverse superconducting systems can be understood from a simple unifying criterion. Additionally, these odd-$\omega$ pair amplitudes have been demonstrated to play multiple roles in determining the properties of these systems. Considering the generality of these phenomena, we believe that many more systems are likely awaiting discovery as odd-$\omega$ superconductors and that as odd-$\omega$ pairing is related to more observable properties it will grow in importance as a means to characterize and understand these systems.

\begin{acknowledgements}
We thank A.~V.~Balatsky, Y.~Gaucher, R. M.~Geilhufe, D.~Kuzmanovski, E.~Langmann, T.~L\"{o}thman, M.~Mashkoori, F.~Parhizgar, B.~Sothmann, and Y.~Tanaka for useful discussions. This work was supported by the Swedish Research Council (Vetenskapsr\aa det) Grant Nos.~2014-3721 and 2018-03488, the Knut and Alice Wallenberg Foundation through the Wallenberg Academy Fellows program, and the European Research Council (ERC) under the European Unions Horizon 2020 research and innovation programme (ERC-2017-StG-757553).
\end{acknowledgements}

\bibliographystyle{andp2012}
\bibliography{Odd_Frequency}

\providecommand{\WileyBibTextsc}{}
\let\textsc\WileyBibTextsc
\providecommand{\othercit}{}
\providecommand{\jr}[1]{#1}
\providecommand{\etal}{~et~al.}


\begin{thebibliography}{[100]}

\bibitem{anderson1959theory}
 \textsc{P.\,W. Anderson} \jr{J. Phys. Chem. Solids} \textbf{11}(1-2), 26--30
  (1959).


\bibitem{ferrell1959knight}
 \textsc{R.\,A. Ferrell} \jr{Phys. Rev. Lett.} \textbf{3}(6), 262 (1959).


\bibitem{anderson1959knight}
 \textsc{P.~Anderson} \jr{Phys. Rev. Lett.} \textbf{3}(7), 325 (1959).


\bibitem{sigrist1991phenomenological}
 \textsc{M.~Sigrist} and  \textsc{K.~Ueda} \jr{Rev. Mod. Phys.} \textbf{63}(2),
  239 (1991).


\bibitem{van1995phase}
 \textsc{D.\,J. Van~Harlingen} \jr{Rev. Mod. Phys.} \textbf{67}(2), 515 (1995).


\bibitem{tsuei2000pairing}
 \textsc{C.~Tsuei} and  \textsc{J.~Kirtley} \jr{Rev. Mod. Phys.}
  \textbf{72}(4), 969 (2000).


\bibitem{balatsky2006impurity}
 \textsc{A.~Balatsky},  \textsc{I.~Vekhter},  and  \textsc{J.\,X. Zhu} \jr{Rev.
  Mod. Phys.} \textbf{78}(2), 373 (2006).


\bibitem{qi2011topological}
 \textsc{X.\,L. Qi} and  \textsc{S.\,C. Zhang} \jr{Rev. Mod. Phys.}
  \textbf{83}(4), 1057 (2011).


\bibitem{eliashberg1960interactions}
 \textsc{G.~Eliashberg} \jr{Sov. Phys. JETP} \textbf{11}(3), 696--702 (1960).


\othercit
\bibitem{abrikosov2012methods}
 \textsc{A.\,A. Abrikosov},  \textsc{L.\,P. Gorkov},  and  \textsc{I.\,E.
  Dzyaloshinski},
Methods of quantum field theory in statistical physics (Courier Corporation,
  2012).


\othercit
\bibitem{mahan2013many}
 \textsc{G.\,D. Mahan},
Many-particle physics (Springer Science \& Business Media, 2013).


\bibitem{scalapino1966strong}
 \textsc{D.~Scalapino},  \textsc{J.~Schrieffer},  and  \textsc{J.~Wilkins}
  \jr{Phys. Rev.} \textbf{148}(1), 263 (1966).


\bibitem{berk1966effect}
 \textsc{N.~Berk} and  \textsc{J.~Schrieffer} \jr{Phys. Rev. Lett.}
  \textbf{17}(8), 433 (1966).


\bibitem{Berezinskii1974}
 \textsc{V.\,L. Berezinskii} \jr{Pis'ma Zh. Eksp. Teor. Fiz.} \textbf{20}, 628
  (1974).


\bibitem{kuzmanovski2017multiple}
 \textsc{D.~Kuzmanovski} and  \textsc{A.\,M. Black-Schaffer} \jr{Phys. Rev. B}
  \textbf{96}(17), 174509 (2017).


\bibitem{geilhufe2018symmetry}
 \textsc{R.\,M. Geilhufe} and  \textsc{A.\,V. Balatsky} \jr{Phys. Rev. B}
  \textbf{97}(2) (2018).


\bibitem{linder2017odd}
 \textsc{J.~Linder} and  \textsc{A.\,V. Balatsky} \jr{arXiv preprint
  1709.03986} (2017).


\bibitem{kirkpatrick_1991_prl}
 \textsc{T.\,R. Kirkpatrick} and  \textsc{D.~Belitz} \jr{Phys. Rev. Lett.}
  \textbf{66}, 1533--1536 (1991).


\bibitem{belitz_1992_prb}
 \textsc{D.~Belitz} and  \textsc{T.\,R. Kirkpatrick} \jr{Phys. Rev. B}
  \textbf{46}, 8393--8408 (1992).


\bibitem{BalatskyPRB1992}
 \textsc{A.~Balatsky} and  \textsc{E.~Abrahams} \jr{Phys. Rev. B} \textbf{45},
  13125 (1992).


\bibitem{coleman_1993_prl}
 \textsc{P.~Coleman},  \textsc{E.~Miranda},  and  \textsc{A.~Tsvelik} \jr{Phys.
  Rev. Lett.} \textbf{70}, 2960--2963 (1993).


\bibitem{coleman_1994_prb}
 \textsc{P.~Coleman},  \textsc{E.~Miranda},  and  \textsc{A.~Tsvelik} \jr{Phys.
  Rev. B} \textbf{49}, 8955--8982 (1994).


\bibitem{coleman_1995_prl}
 \textsc{P.~Coleman},  \textsc{E.~Miranda},  and  \textsc{A.~Tsvelik} \jr{Phys.
  Rev. Lett.} \textbf{74}, 1653--1656 (1995).


\bibitem{heid1995thermodynamic}
 \textsc{R.~Heid} \jr{Z. Phys. B} (1995).


\bibitem{belitz_1999_prb}
 \textsc{D.~Belitz} and  \textsc{T.\,R. Kirkpatrick} \jr{Phys. Rev. B}
  \textbf{60}, 3485--3498 (1999).


\bibitem{solenov2009thermodynamical}
 \textsc{D.~Solenov},  \textsc{I.~Martin},  and  \textsc{D.~Mozyrsky} \jr{Phys.
  Rev. B} \textbf{79}(13), 132502 (2009).


\bibitem{kusunose2011puzzle}
 \textsc{H.~Kusunose},  \textsc{Y.~Fuseya},  and  \textsc{K.~Miyake} \jr{J.
  Phys. Soc. Jpn} \textbf{80}(5), 054702 (2011).


\bibitem{FominovPRB2015}
 \textsc{Y.\,V. Fominov},  \textsc{Y.~Tanaka},  \textsc{Y.~Asano},  and
  \textsc{M.~Eschrig} \jr{Phys. Rev. B} (2015).


\bibitem{bergeret2005odd}
 \textsc{F.\,S. Bergeret},  \textsc{A.\,F. Volkov},  and  \textsc{K.\,B.
  Efetov} \jr{Rev. Mod. Phys.} \textbf{77}, 1321--1373 (2005).


\bibitem{BergeretPRL2001}
 \textsc{F.\,S. Bergeret},  \textsc{A.\,F. Volkov},  and  \textsc{K.\,B.
  Efetov} \jr{Phys. Rev. Lett.} \textbf{86}, 4096--4099 (2001).


\bibitem{halterman2007odd}
 \textsc{K.~Halterman},  \textsc{P.\,H. Barsic},  and  \textsc{O.\,T. Valls}
  \jr{Phys. Rev. Lett.} \textbf{99}(12), 127002 (2007).


\bibitem{yokoyama2007manifestation}
 \textsc{T.~Yokoyama},  \textsc{Y.~Tanaka},  and  \textsc{A.\,A. Golubov}
  \jr{Phys. Rev. B} \textbf{75}, 134510 (2007).


\bibitem{houzet2008ferromagnetic}
 \textsc{M.~Houzet} \jr{Phys. Rev. Lett.} \textbf{101}, 057009 (2008).


\bibitem{EschrigNat2008}
 \textsc{M.~Eschrig} and  \textsc{T.~L{\"o}fwander} \jr{Nat. Phys.} \textbf{4},
  138--143 (2008).


\bibitem{LinderPRB2008}
 \textsc{J.~Linder},  \textsc{T.~Yokoyama},  and  \textsc{A.~Sudb\o{}}
  \jr{Phys. Rev. B} \textbf{77}, 174514 (2008).


\bibitem{crepin2015odd}
 \textsc{F.~Cr\'epin},  \textsc{P.~Burset},  and  \textsc{B.~Trauzettel}
  \jr{Phys. Rev. B} \textbf{92}, 100507 (2015).


\bibitem{YokoyamaPRB2012}
 \textsc{T.~Yokoyama} \jr{Phys. Rev. B} \textbf{86}, 075410 (2012).


\bibitem{Black-SchafferPRB2012}
 \textsc{A.\,M. Black-Schaffer} and  \textsc{A.\,V. Balatsky} \jr{Phys. Rev. B}
  \textbf{86}, 144506 (2012).


\bibitem{Black-SchafferPRB2013}
 \textsc{A.\,M. Black-Schaffer} and  \textsc{A.\,V. Balatsky} \jr{Phys. Rev. B}
  \textbf{87}, 220506 (2013).


\bibitem{TriolaPRB2014}
 \textsc{C.~Triola},  \textsc{E.~Rossi},  and  \textsc{A.\,V. Balatsky}
  \jr{Phys. Rev. B} \textbf{89}, 165309 (2014).


\bibitem{tanaka2007theory}
 \textsc{Y.~Tanaka} and  \textsc{A.\,A. Golubov} \jr{Phys. Rev. Lett.}
  \textbf{98}, 037003 (2007).


\bibitem{TanakaPRB2007}
 \textsc{Y.~Tanaka},  \textsc{Y.~Tanuma},  and  \textsc{A.\,A. Golubov}
  \jr{Phys. Rev. B} \textbf{76}, 054522 (2007).


\bibitem{cayao2017odd}
 \textsc{J.~Cayao} and  \textsc{A.\,M. Black-Schaffer} \jr{Phys. Rev. B}
  \textbf{96}(15), 155426 (2017).


\bibitem{cayao2018odd}
 \textsc{J.~Cayao} and  \textsc{A.\,M. Black-Schaffer} \jr{Phys. Rev. B}
  \textbf{98}, 075425 (2018).


\bibitem{LinderPRL2009}
 \textsc{J.~Linder},  \textsc{T.~Yokoyama},  \textsc{A.~Sudb\o{}},  and
  \textsc{M.~Eschrig} \jr{Phys. Rev. Lett.} \textbf{102}, 107008 (2009).


\bibitem{LinderPRB2010_2}
 \textsc{J.~Linder},  \textsc{A.~Sudb\o{}},  \textsc{T.~Yokoyama},
  \textsc{R.~Grein},  and  \textsc{M.~Eschrig} \jr{Phys. Rev. B} \textbf{81},
  214504 (2010).


\bibitem{TanakaJPSJ2012}
 \textsc{Y.~Tanaka},  \textsc{M.~Sato},  and  \textsc{N.~Nagaosa} \jr{J. Phys.
  Soc. Jpn} \textbf{81}(1), 011013 (2012).


\bibitem{triola2016prl}
 \textsc{C.~Triola},  \textsc{D.\,M. Badiane},  \textsc{A.\,V. Balatsky},  and
  \textsc{E.~Rossi} \jr{Phys. Rev. Lett.} \textbf{116}, 257001 (2016).


\bibitem{triolaprb2016}
 \textsc{C.~Triola} and  \textsc{A.\,V. Balatsky} \jr{Phys. Rev. B}
  \textbf{94}, 094518 (2016).


\bibitem{black2013odd}
 \textsc{A.\,M. Black-Schaffer} and  \textsc{A.\,V. Balatsky} \jr{Phys. Rev. B}
  \textbf{88}, 104514 (2013).


\bibitem{sothmann2014unconventional}
 \textsc{B.~Sothmann},  \textsc{S.~Weiss},  \textsc{M.~Governale},  and
  \textsc{J.~K{\"o}nig} \jr{Phys. Rev. B} \textbf{90}(22), 220501 (2014).


\bibitem{parhizgar_2014_prb}
 \textsc{F.~Parhizgar} and  \textsc{A.\,M. Black-Schaffer} \jr{Phys. Rev. B}
  \textbf{90}, 184517 (2014).


\bibitem{asano2015odd}
 \textsc{Y.~Asano} and  \textsc{A.~Sasaki} \jr{Phys. Rev. B} \textbf{92}(22),
  224508 (2015).


\bibitem{komendova2015experimentally}
 \textsc{L.~Komendov\'a},  \textsc{A.\,V. Balatsky},  and  \textsc{A.\,M.
  Black-Schaffer} \jr{Phys. Rev. B} \textbf{92}, 094517 (2015).


\bibitem{burset2016all}
 \textsc{P.~Burset},  \textsc{B.~Lu},  \textsc{H.~Ebisu},  \textsc{Y.~Asano},
  and  \textsc{Y.~Tanaka} \jr{Phys. Rev. B} \textbf{93}(20), 201402 (2016).


\bibitem{komendova2017odd}
 \textsc{L.~Komendov\'a} and  \textsc{A.\,M. Black-Schaffer} \jr{Phys. Rev.
  Lett.} \textbf{119}, 087001 (2017).


\bibitem{triola2017pair}
 \textsc{C.~Triola} and  \textsc{A.\,V. Balatsky} \jr{Phys. Rev. B}
  \textbf{95}, 224518 (2017).


\bibitem{keidel2018tunable}
 \textsc{F.~Keidel},  \textsc{P.~Burset},  and  \textsc{B.~Trauzettel}
  \jr{Phys. Rev. B} \textbf{97}(7), 075408 (2018).


\bibitem{triola2018odd}
 \textsc{C.~Triola} and  \textsc{A.\,M. Black-Schaffer} \jr{Phys. Rev. B}
  \textbf{97}(6), 064505 (2018).


\bibitem{fleckenstein2018conductance}
 \textsc{C.~Fleckenstein},  \textsc{N.\,T. Ziani},  and  \textsc{B.~Trauzettel}
  \jr{Phys. Rev. B} \textbf{97}(13), 134523 (2018).


\bibitem{asano2018green}
 \textsc{Y.~Asano} and  \textsc{A.\,A. Golubov} \jr{Phys. Rev. B}
  \textbf{97}(21), 214508 (2018).


\bibitem{triola2018oddnw}
 \textsc{C.~Triola} and  \textsc{A.\,M. Black-Schaffer} \jr{arXiv preprint
  1809.09488} (2018).


\bibitem{petrashov1994conductivity}
 \textsc{V.~Petrashov},  \textsc{V.~Antonov},  \textsc{S.~Maksimov},  and
  \textsc{R.\,S. Shaikhaidarov} \jr{JETP Lett.} \textbf{59}(8), 551--555
  (1994).


\bibitem{giroud1998superconducting}
 \textsc{M.~Giroud},  \textsc{H.~Courtois},  \textsc{K.~Hasselbach},
  \textsc{D.~Mailly},  and  \textsc{B.~Pannetier} \jr{Phys. Rev. B}
  \textbf{58}(18), R11872 (1998).


\bibitem{petrashov1999giant}
 \textsc{V.~Petrashov},  \textsc{I.~Sosnin},  \textsc{I.~Cox},
  \textsc{A.~Parsons},  and  \textsc{C.~Troadec} \jr{Phys. Rev. Lett.}
  \textbf{83}(16), 3281 (1999).


\bibitem{aumentado2001mesoscopic}
 \textsc{J.~Aumentado} and  \textsc{V.~Chandrasekhar} \jr{Phys. Rev. B}
  \textbf{64}(5), 054505 (2001).


\bibitem{zhu2010angular}
 \textsc{J.~Zhu},  \textsc{I.\,N. Krivorotov},  \textsc{K.~Halterman},  and
  \textsc{O.\,T. Valls} \jr{Phys. Rev. Lett.} \textbf{105}(20), 207002 (2010).


\bibitem{di2015signature}
 \textsc{A.~Di~Bernardo},  \textsc{S.~Diesch},  \textsc{Y.~Gu},
  \textsc{J.~Linder},  \textsc{G.~Divitini},  \textsc{C.~Ducati},
  \textsc{E.~Scheer},  \textsc{M.\,G. Blamire},  and  \textsc{J.\,W. Robinson}
  \jr{Nat. Commun.} \textbf{6}, 8053 (2015).


\bibitem{di2015intrinsic}
 \textsc{A.~Di~Bernardo},  \textsc{Z.~Salman},  \textsc{X.\,L. Wang},
  \textsc{M.~Amado},  \textsc{M.~Egilmez},  \textsc{M.\,G. Flokstra},
  \textsc{A.~Suter},  \textsc{S.\,L. Lee},  \textsc{J.\,H. Zhao},
  \textsc{T.~Prokscha},  \textsc{E.~Morenzoni},  \textsc{M.\,G. Blamire},
  \textsc{J.~Linder},  and  \textsc{J.\,W.\,A. Robinson} \jr{Phys. Rev. X}
  \textbf{5}, 041021 (2015).


\bibitem{rowell1966electron}
 \textsc{J.~Rowell} and  \textsc{W.~McMillan} \jr{Phys. Rev. Lett.}
  \textbf{16}(11), 453 (1966).


\bibitem{rowell1973tunneling}
 \textsc{J.~Rowell} \jr{Phys. Rev. Lett.} \textbf{30}(5), 167 (1973).


\bibitem{alff1997spatially}
 \textsc{L.~Alff},  \textsc{H.~Takashima},  \textsc{S.~Kashiwaya},
  \textsc{N.~Terada},  \textsc{H.~Ihara},  \textsc{Y.~Tanaka},
  \textsc{M.~Koyanagi},  and  \textsc{K.~Kajimura} \jr{Phys. Rev. B}
  \textbf{55}(22), R14757 (1997).


\bibitem{covington1997observation}
 \textsc{M.~Covington},  \textsc{M.~Aprili},  \textsc{E.~Paraoanu},
  \textsc{L.~Greene},  \textsc{F.~Xu},  \textsc{J.~Zhu},  and  \textsc{C.\,A.
  Mirkin} \jr{Phys. Rev. Lett.} \textbf{79}(2), 277 (1997).


\bibitem{wei1998directional}
 \textsc{J.~Wei},  \textsc{N.\,C. Yeh},  \textsc{D.~Garrigus},  and
  \textsc{M.~Strasik} \jr{Phys. Rev. Lett.} \textbf{81}(12), 2542 (1998).


\bibitem{ebisu2016theory}
 \textsc{H.~Ebisu},  \textsc{B.~Lu},  \textsc{J.~Klinovaja},  and
  \textsc{Y.~Tanaka} \jr{Progress of Theoretical and Experimental Physics}
  \textbf{2016}(8) (2016).


\bibitem{balatsky2018odd}
 \textsc{A.\,V. Balatsky},  \textsc{S.\,S. Pershoguba},  and
  \textsc{C.~Triola} \jr{arXiv preprint 1804.07244} (2018).


\bibitem{maeno1994superconductivity}
 \textsc{Y.~Maeno},  \textsc{H.~Hashimoto},  \textsc{K.~Yoshida},
  \textsc{S.~Nishizaki},  \textsc{T.~Fujita},  \textsc{J.~Bednorz},  and
  \textsc{F.~Lichtenberg} \jr{Nat.} \textbf{372}(6506), 532 (1994).


\bibitem{maeno2012}
 \textsc{Y.~Maeno},  \textsc{S.~Kittaka},  \textsc{T.~Nomura},
  \textsc{S.~Yonezawa},  and  \textsc{K.~Ishida} \jr{J. Phys. Soc. Jpn}
  \textbf{81}(1), 011009 (2012).


\bibitem{hunte2008two}
 \textsc{F.~Hunte},  \textsc{J.~Jaroszynski},  \textsc{A.~Gurevich},
  \textsc{D.~Larbalestier},  \textsc{R.~Jin},  \textsc{A.~Sefat},
  \textsc{M.\,A. McGuire},  \textsc{B.\,C. Sales},  \textsc{D.\,K. Christen},
  and  \textsc{D.~Mandrus} \jr{Nat.} \textbf{453}(7197), 903--905 (2008).


\bibitem{kamihara2008iron}
 \textsc{Y.~Kamihara},  \textsc{T.~Watanabe},  \textsc{M.~Hirano},  and
  \textsc{H.~Hosono} \jr{J. Am. Chem. Soc.} \textbf{130}(11), 3296--3297
  (2008).


\bibitem{ishida2009extent}
 \textsc{K.~Ishida},  \textsc{Y.~Nakai},  and  \textsc{H.~Hosono} \jr{J. Phys.
  Soc. Jpn} \textbf{78}(6), 062001--062001 (2009).


\bibitem{cvetkovic2009multiband}
 \textsc{V.~Cvetkovic} and  \textsc{Z.~Tesanovic} \jr{EPL} \textbf{85}(3),
  37002 (2009).


\bibitem{stewart2011superconductivity}
 \textsc{G.~Stewart} \jr{Rev. Mod. Phys.} \textbf{83}(4), 1589 (2011).


\bibitem{nagamatsu2001superconductivity}
 \textsc{J.~Nagamatsu},  \textsc{N.~Nakagawa},  \textsc{T.~Muranaka},
  \textsc{Y.~Zenitani},  and  \textsc{J.~Akimitsu} \jr{Nat.}
  \textbf{410}(6824), 63--64 (2001).


\bibitem{bouquet2001specific}
 \textsc{F.~Bouquet},  \textsc{R.~Fisher},  \textsc{N.~Phillips},
  \textsc{D.~Hinks},  and  \textsc{J.~Jorgensen} \jr{Phys. Rev. Lett.}
  \textbf{87}(4), 047001 (2001).


\bibitem{brinkman2002multiband}
 \textsc{A.~Brinkman},  \textsc{A.~Golubov},  \textsc{H.~Rogalla},
  \textsc{O.~Dolgov},  \textsc{J.~Kortus},  \textsc{Y.~Kong},
  \textsc{O.~Jepsen},  and  \textsc{O.~Andersen} \jr{Phys. Rev. B}
  \textbf{65}(18), 180517 (2002).


\bibitem{golubov2002specific}
 \textsc{A.~Golubov},  \textsc{J.~Kortus},  \textsc{O.~Dolgov},
  \textsc{O.~Jepsen},  \textsc{Y.~Kong},  \textsc{O.~Andersen},
  \textsc{B.~Gibson},  \textsc{K.~Ahn},  and  \textsc{R.~Kremer} \jr{J. Phys.:
  Condens. Matter} \textbf{14}(6), 1353 (2002).


\bibitem{iavarone2002two}
 \textsc{M.~Iavarone},  \textsc{G.~Karapetrov},  \textsc{A.~Koshelev},
  \textsc{W.~Kwok},  \textsc{G.~Crabtree},  \textsc{D.~Hinks},
  \textsc{W.~Kang},  \textsc{E.\,M. Choi},  \textsc{H.\,J. Kim},
  \textsc{H.\,J. Kim} \etal{} \jr{Phys. Rev. Lett.} \textbf{89}(18), 187002
  (2002).


\bibitem{stewart_prl_1984}
 \textsc{G.\,R. Stewart},  \textsc{Z.~Fisk},  \textsc{J.\,O. Willis},  and
  \textsc{J.\,L. Smith} \jr{Phys. Rev. Lett.} \textbf{52}, 679--682 (1984).


\bibitem{adenwalla1990}
 \textsc{S.~Adenwalla},  \textsc{S.\,W. Lin},  \textsc{Q.\,Z. Ran},
  \textsc{Z.~Zhao},  \textsc{J.\,B. Ketterson},  \textsc{J.\,A. Sauls},
  \textsc{L.~Taillefer},  \textsc{D.\,G. Hinks},  \textsc{M.~Levy},  and
  \textsc{B.\,K. Sarma} \jr{Phys. Rev. Lett.} \textbf{65}, 2298--2301 (1990).


\bibitem{sauls1994}
 \textsc{J.~Sauls} \jr{Adv. Phys.} \textbf{43}(1), 113--141 (1994).


\bibitem{strand_prl_2009}
 \textsc{J.\,D. Strand},  \textsc{D.\,J. Van~Harlingen},  \textsc{J.\,B.
  Kycia},  and  \textsc{W.\,P. Halperin} \jr{Phys. Rev. Lett.} \textbf{103},
  197002 (2009).


\bibitem{strand_science_2010}
 \textsc{J.\,D. Strand},  \textsc{D.\,J. Bahr},  \textsc{D.\,J. Van~Harlingen},
   \textsc{J.\,P. Davis},  \textsc{W.\,J. Gannon},  and  \textsc{W.\,P.
  Halperin} \jr{Science} \textbf{328}(5984), 1368--1369 (2010).


\bibitem{triola2019odd}
 \textsc{C.~Triola} and  \textsc{A.\,M. Black-Schaffer} \jr{arXiv preprint
  1905.00955} (2019).


\bibitem{ramires2016identifying}
 \textsc{A.~Ramires} and  \textsc{M.~Sigrist} \jr{Phys. Rev. B}
  \textbf{94}(10), 104501 (2016).


\bibitem{ishida1998spin}
 \textsc{K.~Ishida},  \textsc{H.~Mukuda},  \textsc{Y.~Kitaoka},
  \textsc{K.~Asayama},  \textsc{Z.~Mao},  \textsc{Y.~Mori},  and
  \textsc{Y.~Maeno} \jr{Nat.} \textbf{396}(6712), 658 (1998).


\bibitem{ishida2015spin}
 \textsc{K.~Ishida},  \textsc{M.~Manago},  \textsc{T.~Yamanaka},
  \textsc{H.~Fukazawa},  \textsc{Z.~Mao},  \textsc{Y.~Maeno},  and
  \textsc{K.~Miyake} \jr{Phys. Rev. B} \textbf{92}(10), 100502 (2015).


\bibitem{duffy2000polarized}
 \textsc{J.~Duffy},  \textsc{S.~Hayden},  \textsc{Y.~Maeno},  \textsc{Z.~Mao},
  \textsc{J.~Kulda},  and  \textsc{G.~McIntyre} \jr{Phys. Rev. Lett.}
  \textbf{85}(25), 5412 (2000).


\bibitem{luke1998time}
 \textsc{G.\,M. Luke},  \textsc{Y.~Fudamoto},  \textsc{K.~Kojima},
  \textsc{M.~Larkin},  \textsc{J.~Merrin},  \textsc{B.~Nachumi},
  \textsc{Y.~Uemura},  \textsc{Y.~Maeno},  \textsc{Z.~Mao},  \textsc{Y.~Mori}
  \etal{} \jr{Nat.} \textbf{394}(6693), 558 (1998).


\bibitem{luke2000unconventional}
 \textsc{G.~Luke},  \textsc{Y.~Fudamoto},  \textsc{K.~Kojima},
  \textsc{M.~Larkin},  \textsc{B.~Nachumi},  \textsc{Y.~Uemura},
  \textsc{J.~Sonier},  \textsc{Y.~Maeno},  \textsc{Z.~Mao},  \textsc{Y.~Mori}
  \etal{} \jr{Physica B} \textbf{289}, 373--376 (2000).


\bibitem{xia_prl_2006}
 \textsc{J.~Xia},  \textsc{Y.~Maeno},  \textsc{P.\,T. Beyersdorf},
  \textsc{M.\,M. Fejer},  and  \textsc{A.~Kapitulnik} \jr{Phys. Rev. Lett.}
  \textbf{97}, 167002 (2006).


\bibitem{nishizaki1999effect}
 \textsc{S.~NishiZaki},  \textsc{Y.~Maeno},  and  \textsc{Z.~Mao} \jr{J. Low
  Temp. Phys.} \textbf{117}(5-6), 1581--1585 (1999).


\bibitem{nishizaki2000changes}
 \textsc{S.~NishiZaki},  \textsc{Y.~Maeno},  and  \textsc{Z.~Mao} \jr{J. Phys.
  Soc. Jpn.} \textbf{69}(2), 572--578 (2000).


\bibitem{deguchi2004gap}
 \textsc{K.~Deguchi},  \textsc{Z.~Mao},  \textsc{H.~Yaguchi},  and
  \textsc{Y.~Maeno} \jr{Phys. Rev. Lett.} \textbf{92}(4), 047002 (2004).


\bibitem{pustogow2019pronounced}
 \textsc{A.~Pustogow},  \textsc{Y.~Luo},  \textsc{A.~Chronister},
  \textsc{Y.\,S. Su},  \textsc{D.~Sokolov},  \textsc{F.~Jerzembeck},
  \textsc{A.~Mackenzie},  \textsc{C.~Hicks},  \textsc{N.~Kikugawa},
  \textsc{S.~Raghu} \etal{} \jr{arXiv preprint 1904.00047} (2019).


\bibitem{mackenzie1996quantum}
 \textsc{A.~Mackenzie},  \textsc{S.~Julian},  \textsc{A.~Diver},
  \textsc{G.~McMullan},  \textsc{M.~Ray},  \textsc{G.~Lonzarich},
  \textsc{Y.~Maeno},  \textsc{S.~Nishizaki},  and  \textsc{T.~Fujita} \jr{Phys.
  Rev. Lett.} \textbf{76}(20), 3786 (1996).


\bibitem{bergemann2000detailed}
 \textsc{C.~Bergemann},  \textsc{S.~Julian},  \textsc{A.~Mackenzie},
  \textsc{S.~NishiZaki},  and  \textsc{Y.~Maeno} \jr{Phys. Rev. Lett.}
  \textbf{84}(12), 2662 (2000).


\bibitem{oguchi1995electronic}
 \textsc{T.~Oguchi} \jr{Phys. Rev. B} \textbf{51}(2), 1385 (1995).


\bibitem{singh1995relationship}
 \textsc{D.\,J. Singh} \jr{Phys. Rev. B} \textbf{52}(2), 1358 (1995).


\bibitem{tou_prl_1996}
 \textsc{H.~Tou},  \textsc{Y.~Kitaoka},  \textsc{K.~Asayama},
  \textsc{N.~Kimura},  \textsc{Y.~\ifmmode\,\bar{O}\else \={O}\fi{}nuki},
  \textsc{E.~Yamamoto},  and  \textsc{K.~Maezawa} \jr{Phys. Rev. Lett.}
  \textbf{77}, 1374--1377 (1996).


\bibitem{schemm_2014}
 \textsc{E.\,R. Schemm},  \textsc{W.\,J. Gannon},  \textsc{C.\,M. Wishne},
  \textsc{W.\,P. Halperin},  and  \textsc{A.~Kapitulnik} \jr{Science}
  \textbf{345}(6193), 190--193 (2014).


\bibitem{yanase_prb_2016}
 \textsc{Y.~Yanase} \jr{Phys. Rev. B} \textbf{94}, 174502 (2016).


\bibitem{yanase_2017_prb}
 \textsc{Y.~Yanase} and  \textsc{K.~Shiozaki} \jr{Phys. Rev. B} \textbf{95},
  224514 (2017).


\bibitem{wang_2017}
 \textsc{Z.~Wang},  \textsc{J.~Berlinsky},  \textsc{G.~Zwicknagl},  and
  \textsc{C.~Kallin} \jr{Phys. Rev. B} \textbf{96}, 174511 (2017).


\bibitem{joynt_rmp_2002}
 \textsc{R.~Joynt} and  \textsc{L.~Taillefer} \jr{Rev. Mod. Phys.} \textbf{74},
  235--294 (2002).


\bibitem{nomoto_prl_2016}
 \textsc{T.~Nomoto} and  \textsc{H.~Ikeda} \jr{Phys. Rev. Lett.} \textbf{117},
  217002 (2016).


\othercit
\bibitem{katsnelson2012graphene}
 \textsc{M.~Katsnelson},
Graphene: carbon in two dimensions (Cambridge university press, 2012).


\bibitem{novoselov2004electric}
 \textsc{K.\,S. Novoselov},  \textsc{A.\,K. Geim},  \textsc{S.\,V. Morozov},
  \textsc{D.~Jiang},  \textsc{Y.~Zhang},  \textsc{S.\,V. Dubonos},
  \textsc{I.\,V. Grigorieva},  and  \textsc{A.\,A. Firsov} \jr{Science}
  \textbf{306}(5696), 666--669 (2004).


\bibitem{neto2009electronic}
 \textsc{A.\,C. Neto},  \textsc{F.~Guinea},  \textsc{N.\,M. Peres},
  \textsc{K.\,S. Novoselov},  and  \textsc{A.\,K. Geim} \jr{Rev. Mod. Phys.}
  \textbf{81}(1), 109 (2009).


\bibitem{vogt2012silicene}
 \textsc{P.~Vogt},  \textsc{P.~De~Padova},  \textsc{C.~Quaresima},
  \textsc{J.~Avila},  \textsc{E.~Frantzeskakis},  \textsc{M.\,C. Asensio},
  \textsc{A.~Resta},  \textsc{B.~Ealet},  and  \textsc{G.~Le~Lay} \jr{Phys.
  Rev. Lett.} \textbf{108}(15), 155501 (2012).


\bibitem{liu2011quantum}
 \textsc{C.\,C. Liu},  \textsc{W.~Feng},  and  \textsc{Y.~Yao} \jr{Phys. Rev.
  Lett.} \textbf{107}(7), 076802 (2011).


\bibitem{davila2014germanene}
 \textsc{M.~D{\'a}vila},  \textsc{L.~Xian},  \textsc{S.~Cahangirov},
  \textsc{A.~Rubio},  and  \textsc{G.~Le~Lay} \jr{New J. Phys.} \textbf{16}(9),
  095002 (2014).


\bibitem{zhu2015epitaxial}
 \textsc{F.\,f. Zhu},  \textsc{W.\,j. Chen},  \textsc{Y.~Xu},  \textsc{C.\,l.
  Gao},  \textsc{D.\,d. Guan},  \textsc{C.\,h. Liu},  \textsc{D.~Qian},
  \textsc{S.\,C. Zhang},  and  \textsc{J.\,f. Jia} \jr{Nat. Mater.}
  \textbf{14}(10), 1020 (2015).


\bibitem{haldane1988model}
 \textsc{F.\,D.\,M. Haldane} \jr{Phys. Rev. Lett.} \textbf{61}(18), 2015
  (1988).


\bibitem{kane2005qshi}
 \textsc{C.\,L. Kane} and  \textsc{E.\,J. Mele} \jr{Phys. Rev. Lett.}
  \textbf{95}, 226801 (2005).


\bibitem{ohta2006controlling}
 \textsc{T.~Ohta},  \textsc{A.~Bostwick},  \textsc{T.~Seyller},
  \textsc{K.~Horn},  and  \textsc{E.~Rotenberg} \jr{Science}
  \textbf{313}(5789), 951--954 (2006).


\bibitem{sarma2011electronic}
 \textsc{S.\,D. Sarma},  \textsc{S.~Adam},  \textsc{E.~Hwang},  and
  \textsc{E.~Rossi} \jr{Rev. Mod. Phys.} \textbf{83}(2), 407 (2011).


\bibitem{mccann2013electronic}
 \textsc{E.~McCann} and  \textsc{M.~Koshino} \jr{Rep. Prog. Phys.}
  \textbf{76}(5), 056503 (2013).


\bibitem{ramasubramaniam2011tunable}
 \textsc{A.~Ramasubramaniam},  \textsc{D.~Naveh},  and  \textsc{E.~Towe}
  \jr{Phys. Rev. B} \textbf{84}(20), 205325 (2011).


\bibitem{zhang2016visualizing}
 \textsc{C.~Zhang},  \textsc{Y.~Chen},  \textsc{J.\,K. Huang},  \textsc{X.~Wu},
   \textsc{L.\,J. Li},  \textsc{W.~Yao},  \textsc{J.~Tersoff},  and
  \textsc{C.\,K. Shih} \jr{Nat. Commun.} \textbf{7}, 10349 (2016).


\bibitem{geim2013van}
 \textsc{A.~Geim} and  \textsc{I.~Grigorieva} \jr{Nat.} \textbf{499}(7459),
  419--425 (2013).


\bibitem{novoselov20162d}
 \textsc{K.~Novoselov},  \textsc{A.~Mishchenko},  \textsc{A.~Carvalho},  and
  \textsc{A.\,C. Neto} \jr{Science} \textbf{353}(6298), aac9439 (2016).


\bibitem{zhang2010crossover}
 \textsc{Y.~Zhang},  \textsc{K.~He},  \textsc{C.\,Z. Chang},  \textsc{C.\,L.
  Song},  \textsc{L.\,L. Wang},  \textsc{X.~Chen},  \textsc{J.\,F. Jia},
  \textsc{Z.~Fang},  \textsc{X.~Dai},  \textsc{W.\,Y. Shan} \etal{} \jr{Nat.
  Phys.} \textbf{6}(8), 584 (2010).


\bibitem{cheng2010landau}
 \textsc{P.~Cheng},  \textsc{C.~Song},  \textsc{T.~Zhang},  \textsc{Y.~Zhang},
  \textsc{Y.~Wang},  \textsc{J.\,F. Jia},  \textsc{J.~Wang},  \textsc{Y.~Wang},
   \textsc{B.\,F. Zhu},  \textsc{X.~Chen} \etal{} \jr{Phys. Rev. Lett.}
  \textbf{105}(7), 076801 (2010).


\bibitem{zhang2011growth}
 \textsc{G.~Zhang},  \textsc{H.~Qin},  \textsc{J.~Chen},  \textsc{X.~He},
  \textsc{L.~Lu},  \textsc{Y.~Li},  and  \textsc{K.~Wu} \jr{Adv. Funct. Mater.}
  \textbf{21}(12), 2351--2355 (2011).


\bibitem{li2010observation}
 \textsc{G.~Li},  \textsc{A.~Luican},  \textsc{J.\,L. Dos~Santos},
  \textsc{A.\,C. Neto},  \textsc{A.~Reina},  \textsc{J.~Kong},  and
  \textsc{E.~Andrei} \jr{Nat. Phys.} \textbf{6}(2), 109 (2010).


\bibitem{bistritzer2010transport}
 \textsc{R.~Bistritzer} and  \textsc{A.\,H. MacDonald} \jr{Phys. Rev. B}
  \textbf{81}(24), 245412 (2010).


\bibitem{dos2012continuum}
 \textsc{J.\,L. dos Santos},  \textsc{N.~Peres},  and  \textsc{A.\,C. Neto}
  \jr{Phys. Rev. B} \textbf{86}(15), 155449 (2012).


\bibitem{cao2018unconventional}
 \textsc{Y.~Cao},  \textsc{V.~Fatemi},  \textsc{S.~Fang},
  \textsc{K.~Watanabe},  \textsc{T.~Taniguchi},  \textsc{E.~Kaxiras},  and
  \textsc{P.~Jarillo-Herrero} \jr{Nat.} \textbf{556}(7699), 43 (2018).


\bibitem{meissner1933neuer}
 \textsc{W.~Meissner} and  \textsc{R.~Ochsenfeld} \jr{Naturwissenschaften}
  \textbf{21}(44), 787--788 (1933).


\othercit
\bibitem{tinkham2004introduction}
 \textsc{M.~Tinkham},
Introduction to superconductivity (Courier Corporation, 2004).


\bibitem{tanaka2005anomalous}
 \textsc{Y.~Tanaka},  \textsc{Y.~Asano},  \textsc{A.\,A. Golubov},  and
  \textsc{S.~Kashiwaya} \jr{Phys. Rev. B} \textbf{72}(14), 140503 (2005).


\bibitem{asano2011unconventional}
 \textsc{Y.~Asano},  \textsc{A.\,A. Golubov},  \textsc{Y.\,V. Fominov},  and
  \textsc{Y.~Tanaka} \jr{Phys. Rev. Lett.} \textbf{107}(8), 087001 (2011).


\bibitem{higashitani2013magnetic}
 \textsc{S.~Higashitani},  \textsc{H.~Takeuchi},  \textsc{S.~Matsuo},
  \textsc{Y.~Nagato},  and  \textsc{K.~Nagai} \jr{Phys. Rev. Lett.}
  \textbf{110}(17), 175301 (2013).


\bibitem{asano2014consequences}
 \textsc{Y.~Asano},  \textsc{Y.\,V. Fominov},  and  \textsc{Y.~Tanaka}
  \jr{Phys. Rev. B} \textbf{90}(9), 094512 (2014).


\bibitem{taylor_prl_2012}
 \textsc{E.~Taylor} and  \textsc{C.~Kallin} \jr{Phys. Rev. Lett.} \textbf{108},
  157001 (2012).


\bibitem{taylor2013anomalous}
 \textsc{E.~Taylor} and  \textsc{C.~Kallin} \jr{J. Phys.: Conf. Ser.}
  \textbf{449}, 012036 (2013).


\bibitem{wysokinski_2012_prl}
 \textsc{K.\,I. Wysoki\ifmmode\,\acute{n}\else \'{n}\fi{}ski},  \textsc{J.\,F.
  Annett},  and  \textsc{B.\,L. Gy\"orffy} \jr{Phys. Rev. Lett.} \textbf{108},
  077004 (2012).


\bibitem{gradhand_2013_prb}
 \textsc{M.~Gradhand},  \textsc{K.\,I. Wysokinski},  \textsc{J.\,F. Annett},
  and  \textsc{B.\,L. Gy\"orffy} \jr{Phys. Rev. B} \textbf{88}, 094504 (2013).


\end{thebibliography}

\end{document}